\documentclass[conference]{IEEEtran}
\IEEEoverridecommandlockouts
\usepackage{subcaption}
\usepackage{multirow}
\usepackage{array}
\usepackage{caption}

\usepackage{cite}
\usepackage{amsmath,amssymb,amsfonts}
\usepackage{algorithmic}
\usepackage{graphicx}
\usepackage{textcomp}
\def\BibTeX{{\rm B\kern-.05em{\sc i\kern-.025em b}\kern-.08em
    T\kern-.1667em\lower.7ex\hbox{E}\kern-.125emX}}

\usepackage{soul}

\ifodd 1
\usepackage{soul, xcolor}

\fi
\begin{document}

\title{Cloaking the Clock: Emulating Clock Skew in Controller Area Networks
%{\footnotesize \textsuperscript{*}Note: Sub-titles are not captured in Xplore and should not be used}
%\thanks{Identify applicable funding agency here. If none, delete this.}
}

\author{Sang Uk Sagong$^{\ast}$, Xuhang Ying$^{\ast}$, Andrew Clark$^{\dagger}$, Linda Bushnell$^{\ast}$, and Radha Poovendran$^{\ast}$\\
$^{\ast}$ Department of Electrical Engineering, University of Washington, Seattle, WA 98195. \\
$^{\dagger}$ Department of Electrical and Computer Engineering, Worcester Polytechnic Institute, Worcester, MA 01609. \\
Email: \{sagong, xhying, lb2, rp3 \}@uw.edu, aclark@wpi.edu
}

%\author{\IEEEauthorblockN{Sang Sagong, Xuhang Ying, Andrew Clark, Radha Poovendran, Linda Poovendran}
%\IEEEauthorblockA{\textit{Department of Electrical Engineering, University of Washington} \\
%\textit{name of organization (of Aff.)}\\
%City, Country \\
%email address}
%\and
%\IEEEauthorblockN{2\textsuperscript{nd} Given Name Surname}
%\IEEEauthorblockA{\textit{dept. name of organization (of Aff.)} \\
%\textit{name of organization (of Aff.)}\\
%City, Country \\
%email address}
%\and
%\IEEEauthorblockN{3\textsuperscript{rd} Given Name Surname}
%\IEEEauthorblockA{\textit{dept. name of organization (of Aff.)} \\
%\textit{name of organization (of Aff.)}\\
%City, Country \\
%email address}
%\and
%\IEEEauthorblockN{4\textsuperscript{th} Given Name Surname}
%\IEEEauthorblockA{\textit{dept. name of organization (of Aff.)} \\
%\textit{name of organization (of Aff.)}\\
%City, Country \\
%email address}
%\and
%\IEEEauthorblockN{5\textsuperscript{th} Given Name Surname}
%\IEEEauthorblockA{\textit{dept. name of organization (of Aff.)} \\
%\textit{name of organization (of Aff.)}\\
%City, Country \\
%email address}
%\and
%\IEEEauthorblockN{6\textsuperscript{th} Given Name Surname}
%\IEEEauthorblockA{\textit{dept. name of organization (of Aff.)} \\
%\textit{name of organization (of Aff.)}\\
%City, Country \\
%email address}
%
%}

\maketitle

%>>> ABSTRACT
\begin{abstract}
\label{sec:abst}
% Key words: automobile, ECU
Automobiles are equipped with Electronic Control Units (ECU) that communicate via  in-vehicle network protocol standards such as Controller Area Network (CAN). These protocols are designed under the assumption that separating in-vehicle communications from external networks is sufficient for protection against cyber attacks. This assumption, however, has been shown to be invalid by recent  attacks in which adversaries were able to infiltrate the in-vehicle network. 
% Recent cyber attacks, in which adversaries were able to infiltrate the in-vehicle network, have shown this assumption to be invalid. 
%These protocols, however, are designed for closed networks that do not communicate with the external environment, and therefore lack cyber security protections such as encryption and message authentication codes. This leaves in-vehicle networks vulnerable to adversaries who compromise one or more ECUs and then inject false messages in order to disrupt safety and performance of the vehicle, for example, by compromising the vehicle's CD player and then spoofing messages from the engine or steering control. 
Motivated by these attacks, intrusion detection systems (IDSs) have been proposed for in-vehicle networks that attempt to detect attacks by making use of device fingerprinting using properties such as clock skew of an ECU.  In this paper, we propose the cloaking attack, an intelligent masquerade attack  in which an adversary modifies the timing of transmitted messages in order to match the clock skew of a targeted ECU. The attack leverages the fact that, while the clock skew is a physical property of each ECU that cannot be changed by the adversary, the estimation of the clock skew by other ECUs is based on network traffic, which, being a  cyber component only, can be modified by an adversary. We implement the proposed cloaking attack and test it  on two IDSs, namely, the current state-of-the-art IDS and a new IDS that we develop based on the widely-used Network Time Protocol (NTP). We implement the cloaking attack on two hardware testbeds, a prototype and a real connected vehicle, and show that it can always deceive both IDSs. We also introduce a new metric called the Maximum Slackness Index   to quantify the effectiveness of the cloaking attack even when the adversary is unable to precisely match the clock skew of the targeted ECU. %We analyze the effectiveness of the cloaking attack when the adversary is unable to precisely match the clock skew of the targeted ECU, and quantify the effectiveness by introducing a maximum slackness index (MSI) parameter. 

\end{abstract}

%\begin{abstract}
%This document is a model and instructions for \LaTeX. This and the IEEEtran.cls file define the components of your paper [title, text, heads, etc.]. *CRITICAL: Do Not Use Symbols, Special Characters, Footnotes, or Math in Paper Title or Abstract.
%\end{abstract}


%>>> SECTION: Introduction
\section{Introduction}
\label{sec:intro}

% Why security becomes an important issue of contemporary car
% P1
Contemporary automobiles are equipped with electronic control units (ECUs) for various functionalities such as vehicle maneuverability, fuel efficiency, and heat, ventilation, and air conditioning.
In order to operate these ECUs properly, the information among ECUs is exchanged via in-vehicle network protocols.
In-vehicle network protocols are based on standards such as the Controller Area Network (CAN), which were developed for closed networks that are isolated from the external environment. Based on the closed network assumption, in-vehicle protocols were not designed for cyber security, and in particular do not provide encryption or message authentication. 

Connected vehicles, however, have an increasingly large and diverse array of outward-facing components in order to provide safety, navigation, and entertainment, which violate the assumption of a closed operating environment. These external interfaces leave connected vehicles vulnerable to attacks in which an adversary compromises one or more outward-facing ECUs (e.g., CD players or cellular radio), gains access to the CAN bus, and then blocks messages sent by other ECUs (denial-of-service (DoS)) or sends spoofed messages that claim to be originated from legitimate ECUs such as steering or engine control (masquerade attack)~\cite{Checkoway:2011:comprehensive}. Such attacks can create spurious alarms to the driver, disable brakes, or cause the vehicle to accelerate uncontrollably, causing serious safety risks to passengers, pedestrians, and other vehicles \cite{Koscher:2010:experimental,Miller:2013:adventure,Miller:2015:remote}.

The cyber vulnerabilities of connected vehicles have motivated development of intrusion detection systems (IDSs) for in-vehicle networks~\cite{Miller:2013:adventure,Shin:2016:finger,Hoppe:2008:security,Muter:2011:entropy}. Due to the lack of cryptographic integrity checks, such IDSs rely on physical invariants of the system. For instance, ECUs typically transmit messages of fixed length and at  fixed frequencies, and the message contents are not expected to vary drastically over time. In \cite{Miller:2013:adventure,Hoppe:2008:security}, mechanisms for detecting DoS attacks by exploiting message periodicity were proposed. Techniques for detecting spoofing attacks based on the low entropy of network traffic were proposed in \cite{Muter:2011:entropy}. As pointed out in \cite{Shin:2016:finger}, however, entropy-based IDSs may be ineffective against intelligent adversaries who mimic the structure and frequency of legitimate traffic. 

%An IDS for detecting such intelligent attacks was proposed in \cite{Shin:2016:finger}. The insight of \cite{Shin:2016:finger} is that different ECUs will have different hardware clocks. Each hardware clock has a distinct frequency  due to variations in the clock's hardware crystal, a property referred to as clock skew~\cite{Moon:1998,Mills:1992:NTP}. The clock skew of each ECU  is unique to that ECU and remains constant over time. Hence, the clock skew provides one possible invariant for uniquely identifying ECUs, and for detecting when one ECU attempts to impersonate another.  Similar clock skew-based fingerprinting methods have been proposed for wireless access points and automotive networks \cite{Jana:2008:on,Kohno:2005:remote,Zander:2008:ICM:1496711.1496726}.

An IDS for detecting such intelligent attacks was proposed in \cite{Shin:2016:finger}, based on the following principles. Each ECU on the CAN bus has a different hardware clock, which has a distinct clock speed due to variations in the clock's hardware crystal, a property referred to as clock skew~\cite{Moon:1998,Mills:1992:NTP}. 
Since all process clocks in an ECU are derived from the hardware clock,  they are affected by the clock skew as a consequence. In particular, the inter-departure times of messages that are periodically transmitted by an ECU will be impacted by its clock skew. If a naive adversary injects the spoofed periodic message from an ECU that is different from the spoofed ECU, the difference in clock skew will change the inter-departure times. 
Hence, an ECU that receives periodically transmitted messages can estimate the clock skew of the transmitting ECU based on the message inter-arrival times. The IDS located at the receiving ECU then detects an attack when a sudden change in estimated clock skew occurs (Fig. \ref{fig:cids_detects_masquerade_attack}). 

%receiving ECU can estimate the clock skew of a t

%Hence, an IDS located at a receiving ECU can estimate the clock skew of a transmitting ECU by observing the inter-arrival times of periodically transmitted messages, and detect an attack when a sudden change in estimated clock skew occurs (Fig. \ref{fig:cids_detects_masquerade_attack}). 

In this paper, we analyze intrusion detection systems that make use of clock skew for detection\footnote{In the remainder of the paper, we will refer to this class of intrusion detection systems as IDS or detectors.}. Our key observation is that an intelligent adversary who realizes that  
the IDS at the receiver ECU computes clock skew using message inter-arrival times can manipulate the inter-departure times to match the clock skew of the targeted ECU and avoid detection. We refer to this intelligent masquerade attack as the \emph{cloaking attack}, as illustrated in Fig.~\ref{fig:cloaking_attack_bypasses_cids}. These results show that, while physical system properties such as clock skew may be helpful in providing security assurances and detecting attacks, intelligent adversaries may still evade detection when physical properties are filtered or mediated through compromised cyber components. We make the following specific contributions:

\begin{figure}[t!]
	\centering
	\begin{subfigure}[h]{0.49\columnwidth} % {0.48\columnwidth}
		\includegraphics[width=\columnwidth]{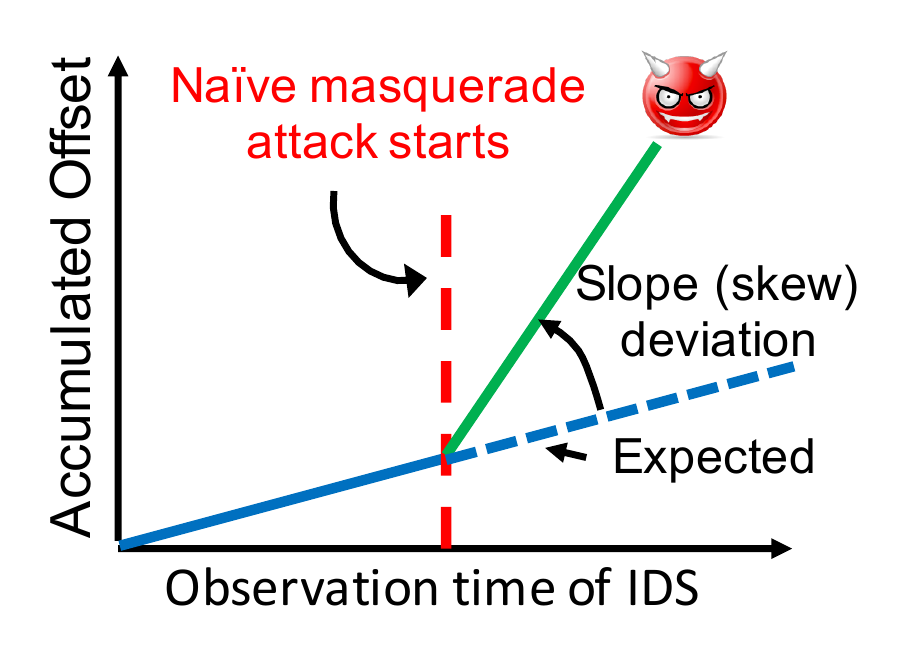}
		\caption{}
		\label{fig:cids_detects_masquerade_attack}
	\end{subfigure}
	\begin{subfigure}[h]{0.49\columnwidth} % {0.48\columnwidth}
		\includegraphics[width=\columnwidth]{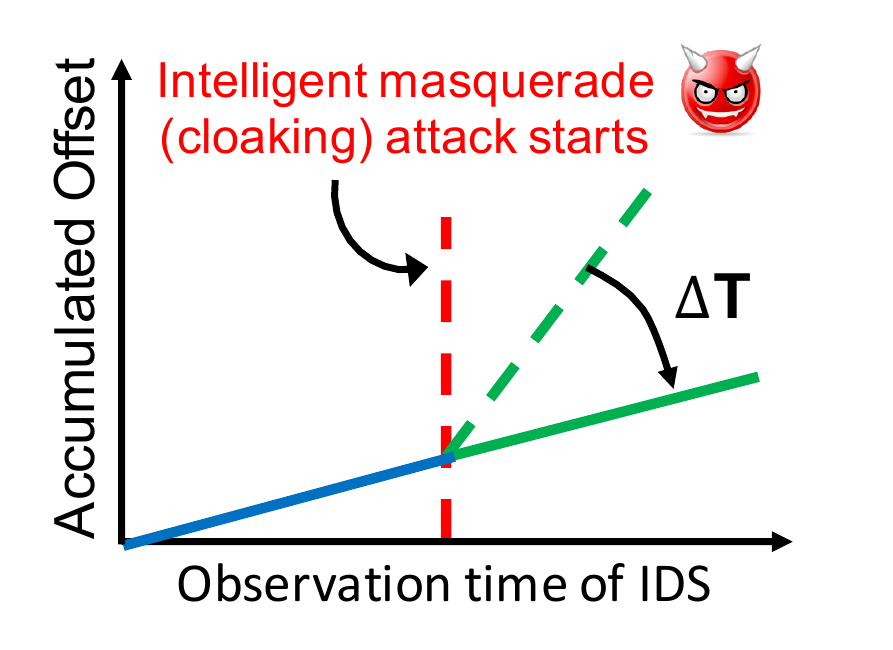}
		\caption{}
		\label{fig:cloaking_attack_bypasses_cids}
	\end{subfigure}
	\caption{Clock skew estimated by the IDS at the receiver. (a) An IDS tracks the clock skew of the transmitter and detects deviations due to naive masquerade attacks. (b) An intelligent masquerading adversary adds a delay $\Delta T$ to message inter-departure times, so as to emulate the clock skew of the targeted ECU and bypass the IDS. }
	\label{fig:masquerading_vs_cloaking}
\end{figure}

\begin{itemize}
\item We propose the cloaking attack, in which an adversary adjusts message timing and cloaks its clock to match the clock of the targeted ECU in order to avoid detection.
\item We analyze the effectiveness of the proposed cloaking attack against two IDSs, including a state-of-the-art IDS and a modified IDS based on the Network Time Protocol (NTP).
\item  We introduce  a new metric called Maximum Slackness Index (MSI) to quantify the effectiveness of an IDS in detecting masquerade attacks.
%The first is the Cho-Shin CIDS proposed in \cite{Shin:2016:finger}. The second is a novel CIDS that we propose based on the Network Timing Protocol (NTP)~\cite{Moon:1998,Mills:1992:NTP,Paxson:1998:CMP:277858.277865}. Our new NTP-based CIDS is motivated by the widespread use of NTP in real-time systems.
\item We evaluate our attack on hardware testbeds, including a CAN bus prototype and a real vehicle (the University of Washington EcoCar). Our hardware evaluations show that the cloaking attack is successful against both IDSs during all hardware trials. 
%\item 
%and the allowed freedom of an attacker in launching the cloaking attack.
%the cloaking attack when there is a deviation between the adversary's clock skew and the clock skew of the targeted ECU. 
We show that the NTP-based IDS has a smaller MSI than the state-of-the-art IDS, and hence is more effective at detecting masquerade attacks. 
%\rev{But $\epsilon\text{-MSI}$ is always non-zero for a CIDS, which indicates the possibility of a cloaking attack.}
\end{itemize}

%
%Hence, the clock skew cannot be used as a signature of ECUs to detect the cloaking attack.
%
%Furthermore, we show how to mount a new bus-off attack using the cloaking attack as a stepping stone.
%

% Organization of paper
The rest of the paper is organized as follows.
Section \ref{sec:cho_shin_cids} explains the adversary model as well as clock-related concepts, and reviews the state-of-the-art IDS.
The NTP-based IDS is introduced in Section \ref{sec:ntp_based_cids}, and the cloaking attack is proposed in Section \ref{sec:cloaking_attack}.
Section \ref{sec:evaluation} presents the experimental results.
%
%We discuss further the significance of controlling the software clock in Section \ref{sec:discuss}.
%
Section \ref{sec:conclude} presents our conclusions and future work.

%>>> SECTION: Overview of Clock-based IDS
\section{Overview of CAN and IDS}
\label{sec:cho_shin_cids}

Below, we review the CAN protocol and needed clock related concepts.
We then present the adversary model, introduce attack scenarios, and review the state-of-the-art IDS 
%proposed by Kyong-Tak Cho and Kang G. Shin 
\cite{Shin:2016:finger}.

\subsection{CAN Background}
\begin{figure}[t!]
	\centering
	\includegraphics[width=.9\columnwidth]{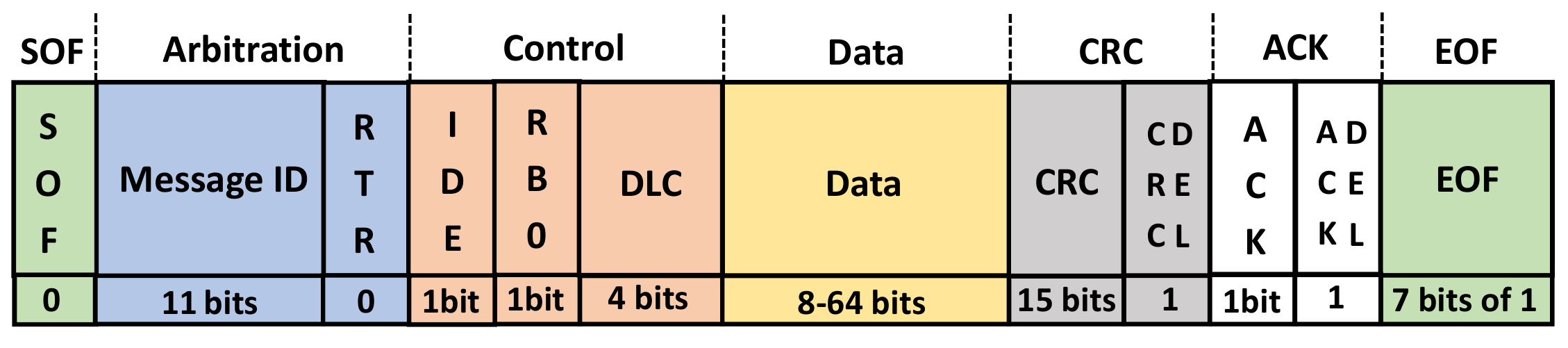}
	\caption{Structure of CAN frame.
		Each frame consists of Start of Frame (SOF) field, Arbitration field, Control field, Data field, CRC field, ACK field, and End of Frame (EOF) field. }
	\label{fig:CAN_DataFrame}
\end{figure}

The CAN protocol \cite{ISO:2015,Bosch:1991} is one of the most widely used in-vehicle networking standards.
%It was developed by Bosch in 1986 to replace the complex wiring harness\cite{Herrewege:2011:CANAuth}. 
%All ECUs are connected to the same and shared bus line, and each ECU accesses the bus via arbitration process which is based on CSMA/CA scheme. Because of high immunity to electrical interference and the ability to self-diagnosis of errors, CAN is used in modern building automation, medical system, and manufacturing \cite{TexasInst:2002:intro}.
CAN is a broadcast bus network, which means that ECUs on the same bus are able to transmit any messages to any ECU and observe all ongoing transmissions.
The CAN frame structure is illustrated in Fig.~\ref{fig:CAN_DataFrame}.
It does not include encryption, authentication, or timestamps. %either encryption or authentication mechanisms. 
%Besides no timestamps are embedded into CAN frames.
%Furthermore, time instants of each ECU are based on its own local clock, and there is no clock synchronization among ECUs. 
%\cmt{Need to describe the arbitration process here. }

The CAN bus acts as a logical AND gate, that is, if two ECUs transmit simultaneously, the message with a smaller ID (higher priority) will be transmitted, 
 through a process known as arbitration.
%\rev{Specifically,} when an ECU transmits the ID value of its message \rev{one bit at a time (starting from the most significant bit)} but observes a different bit \rev{from the bus}, it will stop transmitting and re-transmit at a later time. 
For example, if messages 0x100 and 0x010 are transmitted simultaneously, the ECU that attempts to transmit its message ID 0x100 one bit at a time (starting from the most significant bit) will observe a $0$ bit on the CAN bus although it had transmitted a $1$, recognize that another ECU is transmitting a higher priority message, and stop its transmission.

\subsection{Clock-Related Concepts}
In this section, we follow the Network Time Protocol (NTP) definitions of clocks  \cite{Moon:1998,Mills:1992:NTP,Paxson:1998:CMP:277858.277865}.
Let us first define ${C_{true}}$ as the ``true'' clock that  runs at a constant rate, i.e., ${C_{true}}(t)=t$.
Let $C_A(t)$ denote the time kept by clock $A$. 
The \textit{clock offset} of $C_A$, denoted as $O_A(t)$, is the difference between the time reported by $C_A$ and the ``true'' time, i.e.,
\begin{equation}
O_A(t) = C_A(t) - C_{true}(t).\label{eq:NTP_def_offset}
\end{equation}
The \textit{frequency} of $C_A$ at time $t$ is given by $C'_A(t)$. 
The \textit{clock skew} of $C_A$, denoted as $S_A(t)$, is the difference in the frequencies (or first derivatives) of $C_A$ and $C_{true}$, i.e.,
\begin{equation}
S_A(t) = C_A'(t) - C_{true}'(t).
\end{equation}
A positive clock skew means that $C_A$ runs faster than the true clock, while a negative clock skew implies that $C_{A}$ runs slower than the true clock.
The unit of skew is microseconds per second ($\mu$s/s) or parts per million (ppm). 
For example, if $C_A$ is faster by $5\mu$s every $10$ms according to $C_{true}$, then its skew relative to $C_{true}$ is $500$ppm.

In a vehicle, ECUs  typically have constant clock skews \cite{Shin:2016:finger}. 
Suppose that $C_A$ has a constant skew $S_A$.
 If $\Delta t$ is the time duration measured by $C_{true}$, then the amount of time that has passed according to $C_A$ is 
%\begin{equation}
$\Delta t_A = (1+S_A) \cdot \Delta t$,
%\end{equation} 
and $\Delta t = \Delta t_{A}/(1+S_{A})$. 
Similarly, if there is a second non-true clock $B$ with a constant skew $S_B$ that reports a time duration of $\Delta t_B$, we have $\Delta t_B = (1+S_B)\cdot \Delta t$.
Then the skew of $C_B$ relative to $C_A$, denoted as $S_{BA}$, is given by
\begin{equation}
S_{BA} = \frac{\Delta t_B- \Delta t_A}{\Delta t_A} = \frac{S_B - S_A}{1+S_A} \label{eq:relative_skew}
\end{equation}
and the relationship between $S_{BA}$ and $S_{AB}$, i.e., the skew of $C_A$ relative to $C_B$, is given by
\begin{equation}
S_{AB} = \frac{-S_{BA}}{1+S_{BA}}. \label{eq:relative_skew_conversion}
\end{equation}

When such a ``true'' clock does not exist, a non-true clock is  chosen as the \textit{reference} clock.
Then \textit{relative offset} and \textit{relative skew} are defined for other clocks with respect to the reference clock.
%\footnote{For convenience, we refer to relative offset and relative skew w.r.t. the reference clock as offset and skew, respectively, unless indicated otherwise.}.
Two clocks are said to be \textit{synchronized} at a particular moment if both relative offset and relative skew are zero. 

\subsection{Adversary Model and Attack Scenarios}
  Adversaries can compromise one or more ECUs in a vehicle physically or remotely by exploiting various attack surfaces \cite{Checkoway:2011:comprehensive}. 
As in \cite{Shin:2016:finger}, we consider two types of attackers with different capabilities: 1) \textit{weak attacker}, who is assumed to be able to suspend the transmission of messages of the weakly compromised ECU, but cannot inject any messages, and 2) \textit{strong attacker}, who is assumed to be able to suspend messages of the fully compromised ECU and inject arbitrary attack messages. 

The two types of attackers naturally lead to three attack scenarios: \textit{suspension}, \textit{fabrication}, and \textit{masquerade} attacks. 
In a suspension attack, a weakly compromised ECU is prevented from transmitting certain messages, whereas in a fabrication attack, a fully compromised ECU injects fabricated messages with legitimate IDs. 
%It has been shown in \cmt{[cite]} that both suspension and fabrication attacks can cause severe consequences to the vehicle. 
Since most in-vehicle CAN messages are periodic, the above two attacks would significantly change the frequency of certain messages, and thus can be easily detected by state-of-the-art IDSs \cite{Hoppe:2008:security,Muter:2011:entropy,Muter:2010:structured,Miller:2013:adventure}.

\begin{figure}[t]
	\centering
	\includegraphics[width=0.95\columnwidth]{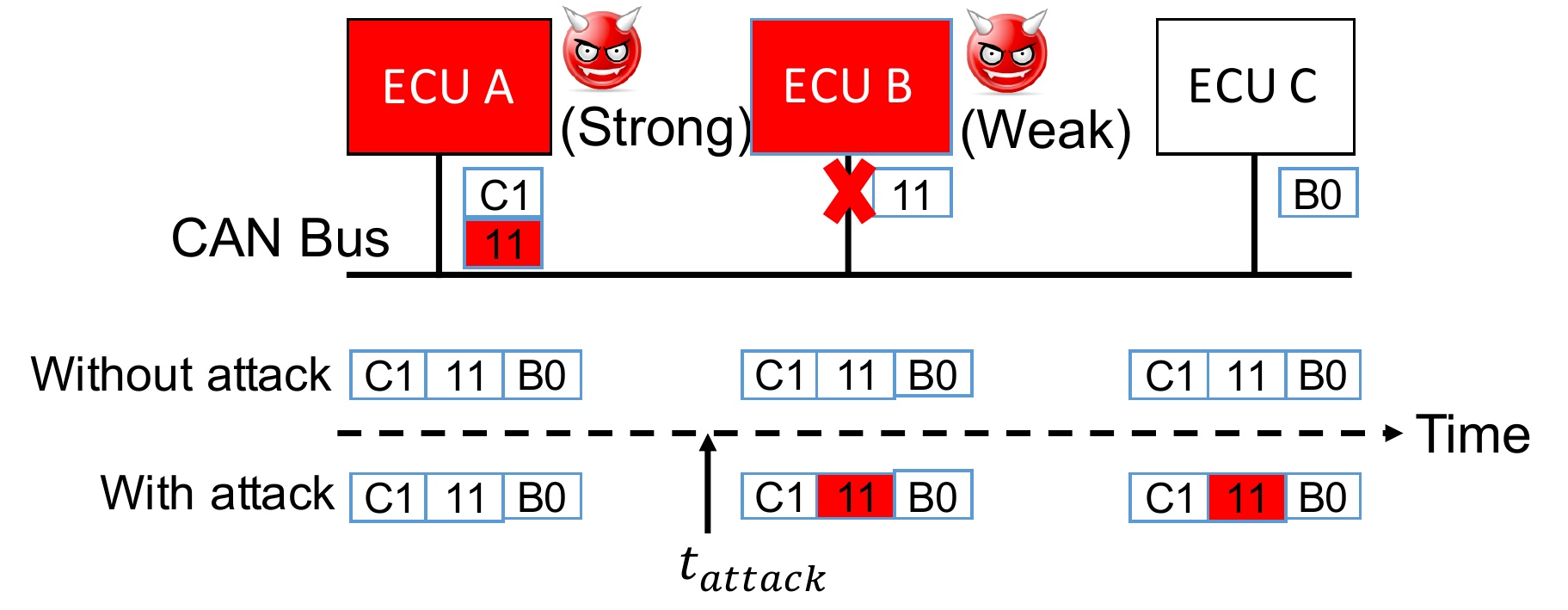}
	\caption{Illustration of masquerade attack. 
		In this example, ECU A is fully compromised by the strong attacker, and ECU B is weakly compromised by the weak attacker.
		Before the attack, ECU B transmits message 0x11 every $T$ sec. 
		At $t=t_{attack}$, the weak attacker suspends ECU B's transmission of message 0x11, and the strong attacker starts fabricating and injecting spoofed messages with ID=0x11 every $T$ sec.
	}
	\label{fig:masquerade_attack_model}
\end{figure} 

Masquerade attacks combine suspension and fabrication attacks. In a masquerade attack, two ECUs A and B are compromised by strong and weak attackers respectively (Fig.~\ref{fig:masquerade_attack_model}). The goal of the attack is to impersonate ECU B by injecting periodic messages with spoofed IDs. During the attack, the weak attacker who has compromised ECU B suspends certain messages from ECU B, while the strong attacker uses the fully compromised ECU A to inject messages claiming to originate from ECU B. It has been shown  that the masquerade attack can potentially cause severe problems to the vehicle \cite{Miller:2015:remote, Wired:Hackers}. Although the previously mentioned IDSs actively monitor the bus traffic, the masquerade attack does not change the frequency of the spoofed message, and thus is more difficult to detect than the suspension and fabrication attacks. 

%A masquerade attack, on the other hand, requires two ECUs to be compromised, one as a weak attacker, and the other as a strong attacker, as illustrated in Fig.~\ref{fig:masquerade_attack_model}.
%The strong attacker is utilized to overcome the weak attacker's inability of injecting messages.
%The adversary's objective is to masquerade a legitimate ECU by injecting periodic messages with spoofed ID at its original frequency, so as to manipulate the receiving ECU while covering the fact that the ECUs has been compromised.
%It has been shown in many studies that the masquerade attack can potentially cause severe problems to the vehicle \revs{\cite{Miller:2015:remote, Wired:Hackers}}.
%Although the previously mentioned IDS's actively monitor the bus traffic, the masquerade attack does not change the frequency of the spoofed message, and thus is more difficult to detect, as compared to the suspension and fabrication attacks. 
%Therefore, we study the masquerade attack in this work, and the recently proposed clock-based detection scheme.

\subsection{Clock-Based Detection}
In-vehicle ECUs operate according to their local clocks with distinct skews, which can be exploited for fingerprinting.
Methods proposed in \cite{Jana:2008:on,Kohno:2005:remote,Zander:2008:ICM:1496711.1496726}, however, are not applicable to CAN since there are no transmit timestamps in CAN messages. 
%However, since there are no transmission timestamps in CAN messages, methods proposed in \cite{Jana:2008:on,Kohno:2005:remote,Zander:2008:ICM:1496711.1496726} are not applicable to CAN.
To bypass this issue, the state-of-the-art IDS in \cite{Shin:2016:finger} uses message periodicity to extract and estimate the transmitters' clock skews for identification. We now review the IDS of \cite{Shin:2016:finger}.

%Cho and Shin considered the use of message periodicity to extract and estimate the transmitters' clock skews for identification \cite{Shin:2016:finger}. 
%In the rest of this section, we will briefly review the state-of-the-art IDS proposed by Cho and Shin.

\begin{figure}[t!]
	\centering
	\includegraphics[width=0.9\columnwidth]{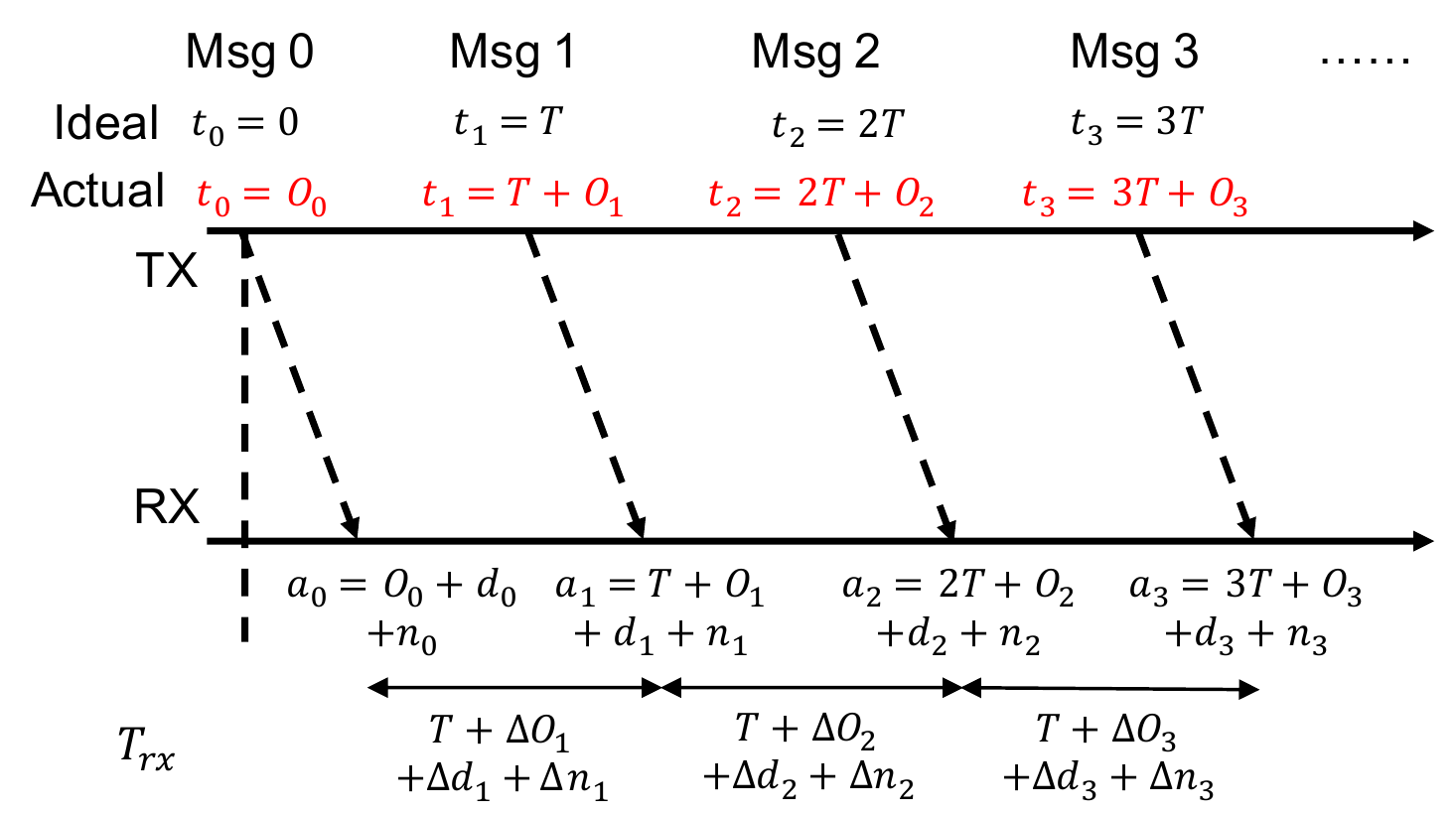}
	\caption{Timing analysis of message arrivals in CAN.}
	\label{fig:offset_model}
\end{figure} 

\subsubsection{Timing model for CAN}
Fig.~\ref{fig:offset_model} illustrates the timing of a periodic message from the perspective of a receiving ECU R. 
Since only R's timestamps are available, we consider its clock as the reference, and refer to the relative offset and relative skew of the transmitter's clock as offset and skew, respectively.
 
Suppose that the transmitter transmits a message every $T$ sec according to its local clock. 
%Without loss of generality, we assume zero offset between the two clocks when the very first message, labeled as $0$, is transmitted.
In the ideal case where the two clocks are synchronized, message $i$ will be transmitted at $t_i = iT$ in R's clock\footnote{Strictly speaking, $t_i$ is the time when the transmitter puts the first bit of message $i$ into the outgoing buffer. }.
Due to clock skew, however, the actual transmission time  is $t_i=iT + O_i$ in R's clock, where $O_i$ is the \textit{accumulated offset} since message $0$. 
After a network delay of $d_i$ (due to message transmission, propagation and reception), the message arrives at the incoming buffer of $R$, and has a timestamp $a_i = iT + O_i + d_i + n_i$, where $n_i$ is zero-mean noise introduced by R's timestamp quantization process \cite{Zander:2008:ICM:1496711.1496726}.
Denote the inter-arrival time between message $(i-1)$ and $i$ as $T_{rx,i}$, which is given by 
\begin{align*}
T_{rx,i} &= T + (O_i - O_{i-1}) + (d_{i} - d_{i-1}) + (n_i-n_{i-1}) \\
&= T + \Delta O_i + \Delta d_i + \Delta n_i,
\end{align*}
where $\Delta O_i$ is the offset in period $i$, and $\Delta d_i$ and $\Delta n_i$ are the differences in network delay and quantization noise, respectively, between periods $i$ and $(i-1)$. 
Since messages with the same ID typically have the same length, it is reasonable to assume $\mathbb{E}[\Delta d_i]=0$. 
%, otherwise, R can easily adjust the timestamps to ensure consistency. 
Since $\mathbb{E}[n_i]=0$ and hence $\mathbb{E}[\Delta n_i]=0$, we have $\mathbb{E}[T_{rx,i}]=T + \mathbb{E}[\Delta O_i]$.

\subsubsection{Clock Skew Detector }
To estimate clock skew,  incoming messages are processed in batches of size $N$ (e.g., $20$), and computes the ``average offset'' in the $k$-th batch,
\begin{equation}
O_{avg}[k] = \frac{1}{N-1} \sum_{i=2}^N\left[a_i - (a_1 + (i-1)\mu_{T}[k-1]) \right], \label{eq:cho_shin_avg_offset}
\end{equation}
where $\mu_{T}[k-1]$ is the average inter-arrival time of the previous batch, and the quantity in the square brackets is the difference between the measured arrival time and the estimated arrival time for the $i$-th message.

When an average offset value is computed from the current batch, its \textit{absolute} value is added to the accumulated offset,  
\begin{equation}
O_{acc}[k]=O_{acc}[k-1]+|O_{avg}[k]|.
\label{eq:cho_shin_acc_offset}
\end{equation}
%which is observed to have linear relationship.
It is then modeled as $O_{acc}[k]=S[k]\cdot t[k] + e[k]$, where $S[k]$ is the regression parameter, $t[k]$ the elapsed time, and $e[k]$ the identification error. 
To estimate the unknown parameter $S$, the Recursive Least Squares (RLS) algorithm is adopted, which minimizes the sum  of squares of the modeling errors \cite{Haykin:2011}.

In a naive masquerade attack, the impersonating ECU has a clock skew different from the targeted ECU's, which would lead to significant identification errors. 
Hence, the identification error is considered as an indicator of whether an attack is taking place.
The IDS tracks the normal clock behavior for messages with the target ID by tracking the mean and variance of the errors (denoted as $e$), $\mu_e$ and $\sigma_e^2$.
To be robust against noise, $\mu_e$ and $\sigma_e^2$ are updated only if $|(e-\mu_{e})/(\sigma_{e})| < \gamma$, where $\gamma$ is a given update threshold. 
For detection, the Cumulative Sum (CUSUM) method, which derives the cumulative sums of deviations from the norm behavior \cite{Basseville:1993}, is implemented. 
Letting $\theta_{e} = \frac{e-\mu_{e}}{\sigma_{e}}$, the upper and lower control limits $L^+$ and $L^-$ are updated for each new error sample as: $$L^+ = \max(0, L^+ + \theta_{e} - \kappa), L^- = \max(0, L^- -\theta_{e} - \kappa),$$ where $\kappa$ is a sensitivity parameter. 
If either control limit exceeds a detection threshold $\Gamma$, a sudden shift is detected, and the IDS declares an intrusion.
The values of $\gamma$, $\kappa$, and $\Gamma$ chosen in \cite{Shin:2016:finger} are 3, 5, and 5, respectively.
A more detailed workflow of the state-of-the-art IDS is provided in Appendix~\ref{appendix:workflow}.

%Since the clock skew is shown to be a constant, as will be discussed in Section \ref{sec:ntp_based_cids}, we derive clock offset based on NTP definition that $E[\Delta O_i]$ is assumed to be a constant \cite{Kohno:2005:remote,Moon:1998,Paxson:1998:CMP:277858.277865}.

%
%In this paper, we consider two different cases of suspending messages depending on ECU B's capability. Let's assume that ECU B transits multiple legitimate messages.
%
%\begin{itemize}
%	\item ECU B suspends all of its message, and all of messages are spoofed by ECU A.
%	\item ECU B cannot suspend all of its messages. That is, part of its messages are suspended and the other messages are transmitted after the masquerade attack.
%\end{itemize}

%
%Since the local clock has its own clock skew, each node in the network can be identified using clock skew as shown in \cite{Jana:2008:on,Zander:2008:ICM:1496711.1496726,Amsthm15}.
%

\subsubsection{Correlation Detector}
%As an optional feature, a correlation detector may be employed. %if the message tracked by the IDS has a sibling message that is periodically transmitted by the same transmitter.
It is pointed out in \cite{Shin:2016:finger} that if two messages are from the same transmitter, their average offsets are likely to be equivalent and show high correlation (i.e., the correlation coefficient $\rho$ is close to $1$), whereas two messages from different ECUs would have low correlation.
Hence, the correlation detector keeps track of the correlation of two highly correlated messages, and declares a masquerade attack if $\rho$ is less than a detection threshold (e.g., 0.8).
As a result, in cases where the impersonating ECU happens to have a similar clock skew with the targeted ECU, the masquerade attack may bypass the clock skew detector, but would still be detected by the correlation detector. 
It is important to note that the clock skew detector applies to any periodic message, but the correlation detector is only applicable to a pair of messages with highly correlated average offsets. 

We will use analytical and experimental analyses in Sections \ref{sec:cloaking_attack} and \ref{sec:evaluation} to show that  not all pairs of messages from the same transmitter show high correlation.
Specifically, we find that high correlation is more likely to exist between two messages that are \textit{consecutively transmitted} by the same ECU and also \textit{consecutively received} by the receiver.

%>>> SECTION: NTP-based CIDS
\section{NTP-based IDS}
\label{sec:ntp_based_cids}
In this section, we present a modified detector that computes clock offsets and skews according to the NTP specifications, which is referred to as the NTP-based IDS. The main difference between the two detectors is the computation of the clock skew, described below.

The motivation for our NTP-based IDS is two-fold. First, we note that the  the metric in Eq.~(\ref{eq:cho_shin_avg_offset}) is not consistent with the NTP definition in Eq.~(\ref{eq:NTP_def_offset}), since it does not calculate the time difference between the transmitter's clock and the reference clock. In addition, it is assumed that $O_i$ is a random variable and $\mathbb{E}[\Delta O_i]=0$, which implies that $\mathbb{E}[O_i]=\mathbb{E}[O_j]$ for $i\neq j$, which does not hold in general since offsets accumulate over time (e.g., if $i\gg j$, $\mathbb{E}[O_i] \gg \mathbb{E}[O_j]$).
%clocks have different skews. 
Our second motivation is the widespread use and acceptance of NTP as a timing mechanism for real-time systems, which raises the question of whether NTP can be used for intrusion detection as well. 
%But this assumption may not hold in practice: if we let $i\rightarrow \infty$, we would expect $\mathbb{E}[O_i] \gg \mathbb{E}[O_1]$, because offsets, although small in one period, would accumulate over time. 
%For example, if a clock provides an accurate reading at $t=0$ but is slower than the true clock by approximately $1$ms every hour, it will be slower by $3$ms after three hours.
%In Section \ref{sec:ntp_based_cids}, we will present our timing model, and calculate average offset according to NTP definitions.

\subsection{Clock Skew Estimation in NTP}
In the NTP-based IDS, the accumulated offset up to message $i$ in Fig.~\ref{fig:offset_model} is modeled as a random variable, $O_i = i\cdot O + \epsilon_i$, where $O$ is the offset in each period $T$ given the constant clock skew, and $\epsilon_i$ is the offset deviation due to ECU jitters.
We assume that $\mathbb{E}[\epsilon_i]=0$ and $\{\epsilon_i\}$ are independent of each other. 
Hence, the expected accumulated offset is $\mathbb{E}[O_i]=i\cdot O$, which increases linearly as more messages are transmitted. 

Consider two consecutively received messages with timestamps $a_{i-1}$ and $a_i$.
From the receiver's perspective, the message period is $T$ in the transmitter's clock, which corresponds to $T_{rx,i}=a_i - a_{i-1}$ (i.e., the observed period) in the receiver's clock. 
By the definition in Eq.~(\ref{eq:NTP_def_offset}), the observed offset is
\begin{equation*}
\hat{O}_i = T - (a_i - a_{i-1})=- (O + \Delta \epsilon_i + \Delta d_i + \Delta n_i), 
\end{equation*}
where $\Delta \epsilon_i = \epsilon_i-\epsilon_{i-1}$.
A batch of $N$ messages is used to compute the average offset of the $k$-th batch $O_{avg}[k]$, i.e.,
\begin{equation}
O_{avg}[k]=\frac{1}{N}\sum_{i=1}^N \hat{O}_i = \frac{1}{N}\sum_{i=1}^N [T-(a_i - a_{i-1})] = T - \frac{a_N - a_0}{N},
\label{eq:ntp_avg_offset}
\end{equation}
where $a_0$ is the timestamp of the last message in the previous batch. 
%It is easy to see that $\hat{O}$ is an unbiased estimator for $O$, since $\mathbb{E}[\hat{O}] = O$.
The accumulated offset up to the last message of the $k$-th batch is given by: 
\begin{equation}
O_{acc}[k] = O_{acc}[k-1] + N \cdot O_{avg}[k].
\label{eq:ntp_acc_offset}
\end{equation}
Note that the original value of $O_{avg}[k]$ is used, instead of the absolute value as in the state-of-the-art IDS.
The other components of the NTP-based IDS remains the same as the state-of-the-art IDS. 
More details are available in Appendix~\ref{appendix:workflow}.

%\cmt{What else is different between NTP-based IDS and state-of-the-art IDS? The latest $n$ error samples are used for estimating $\mu_e$ and $\sigma_e$?}
%\revs{In the state-of-the-art IDS, $\mu_T$ is computed every batch in the receiver's local clock. However, in the NTP-based IDS, the period of the message $T$ is a fixed period in the transmitter's local clock which follows the NTP definition. Hence, the clock skew is indeed estimated by the NTP-based IDS. In the state-of-the-art IDS, $\mu_e$ and $\sigma_e$ are updated using all previous identification error samples. However, most $n$ recent identification error samples in the NTP-based IDS, hence, the NTP-based IDS can effectively allocate hardware resources such as memory which is also more practical with respect to implementation. }

\subsection{Estimation Consistency}
As a physical property of an ECU, clock skew is considered to be stable over time, and thus the estimated values should be  \textit{consistent}, across 1) different batch sizes used by an IDS, 2) different portions of the same trace, and 3) different traces of the same ECU.
Hence, we use the Toyota Camry dataset \cite{Ruth:2012:Accuracy} that was used in \cite{Shin:2016:finger} to compare the NTP-based IDS against the state-of-the-art IDS in terms of estimation consistency.

%This dataset contains CAN traffic data from a Toyota Camry 2010, collected in $14$ different experiments; 
%\footnote{\revs{Ruth \emph{et. al.} provide CAN traffic data logged from a Toyota Camry 2010.
		%and the data is logged through a Gryphon S3 and Hercules software.
%		They ran 14 independent experiments, and each experiment is about 4 minutes long. In Toyota Camry 2010, there are 42 distinct messages transmitted on the CAN bus.
		%In \cite{Shin:2016:finger}, the transmitter of each message is identified using the method discussed in \cite{Natale:2012} which can group messages according to their sources.
%}}

\begin{figure}
	\centering
	\begin{subfigure}[b]{0.49\columnwidth}
	\includegraphics[width=\columnwidth]{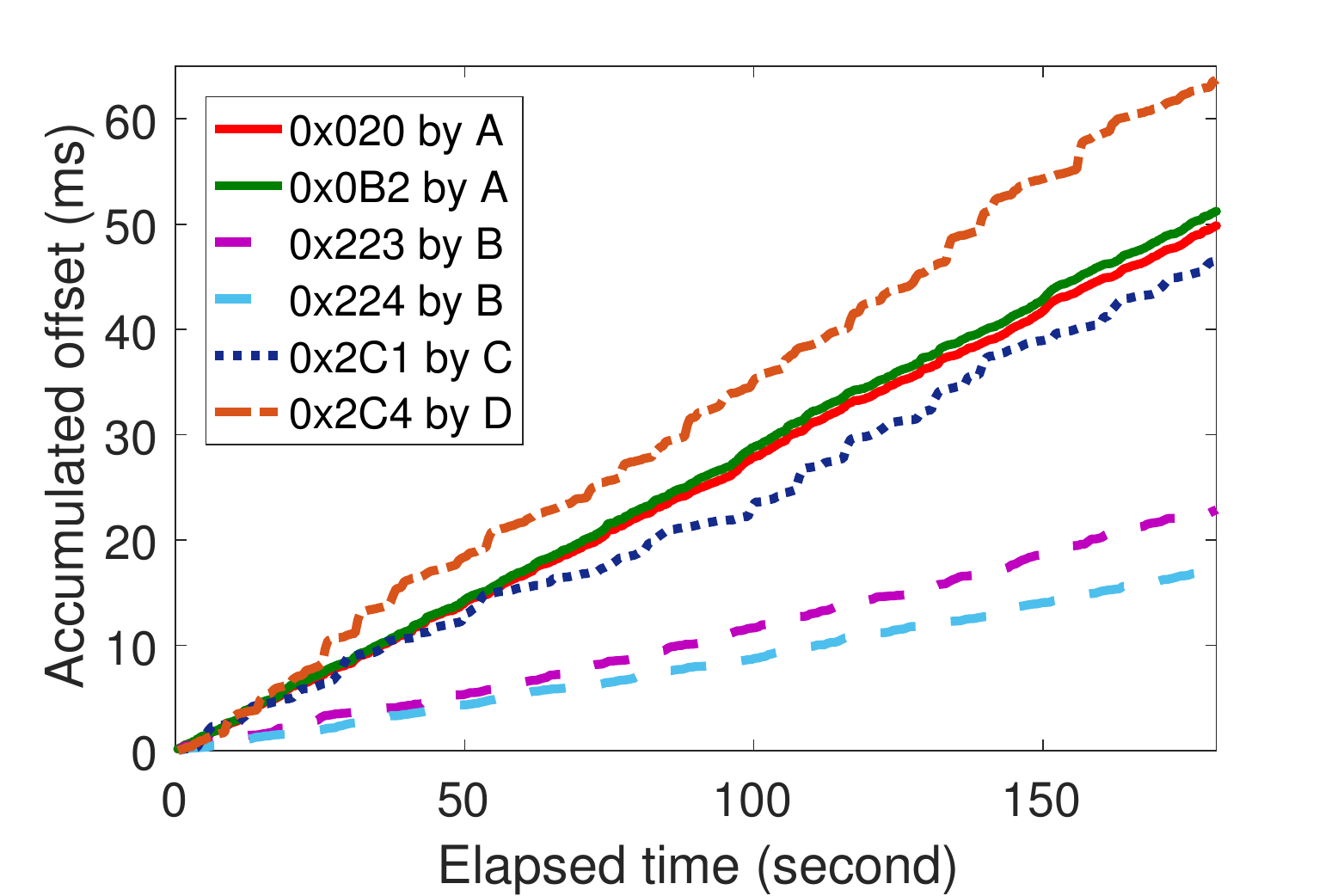}
	\caption{State-of-the-art IDS, $N=20$}
	\end{subfigure}
	\begin{subfigure}[b]{0.49\columnwidth}
	\includegraphics[width=\columnwidth]{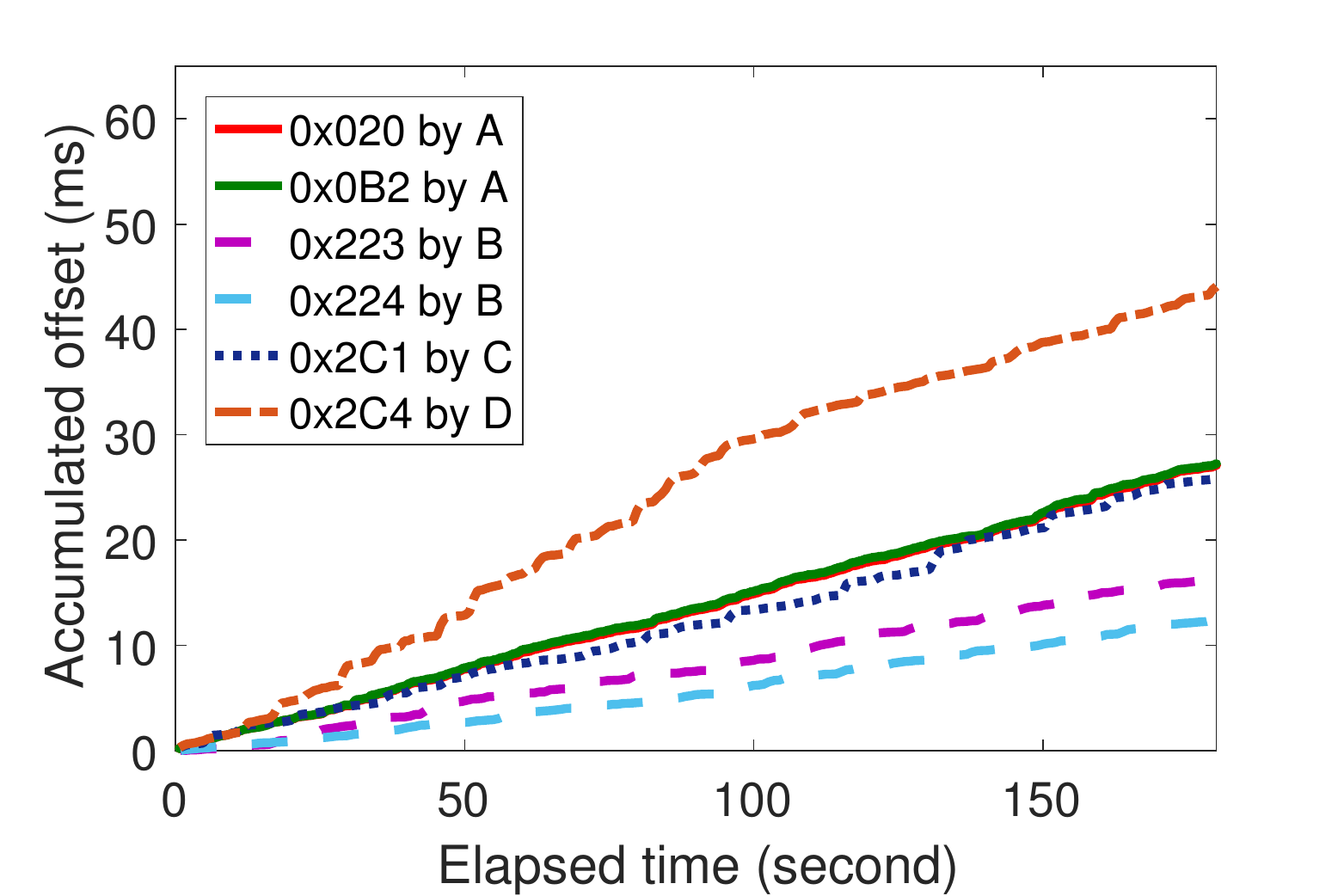}
	\caption{State-of-the-art IDS, $N=30$}		
	\end{subfigure}
	\\
	\begin{subfigure}[b]{0.49\columnwidth}
		\includegraphics[width=\columnwidth]{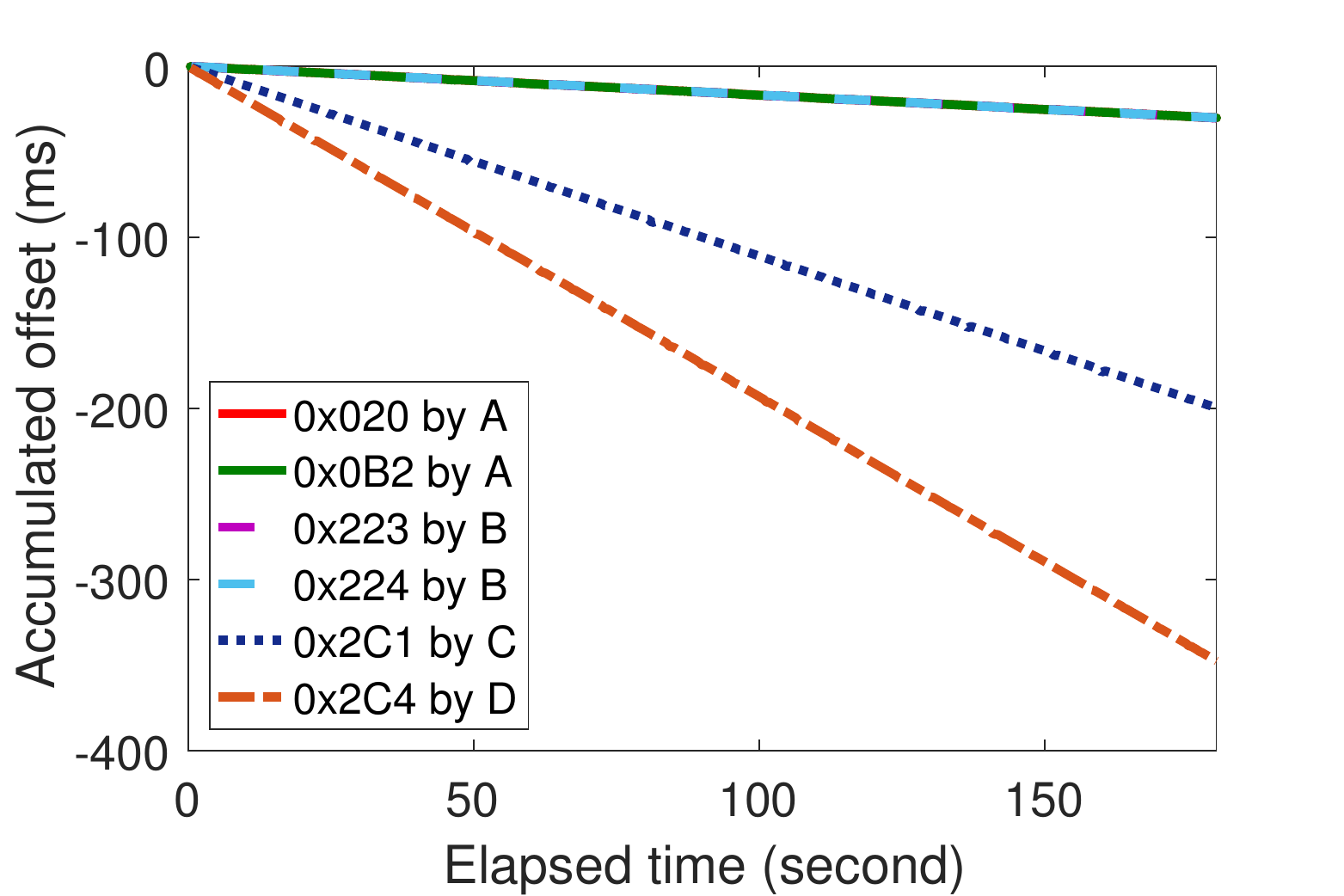}
		\caption{NTP-based IDS, $N=20$}
	\end{subfigure}
	\begin{subfigure}[b]{0.49\columnwidth}
		\includegraphics[width=\columnwidth]{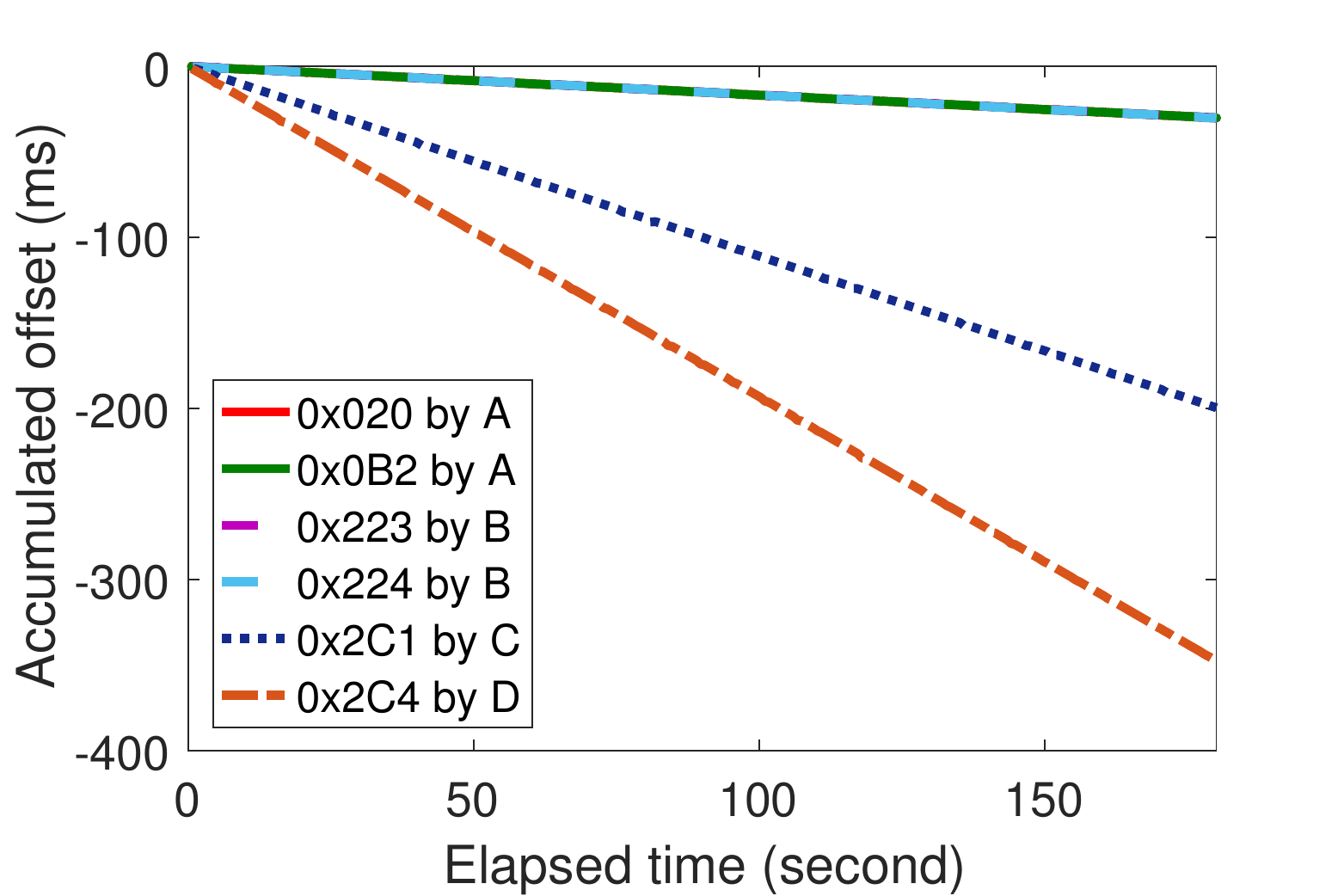}
		\caption{NTP-based IDS, $N=30$}		
	\end{subfigure}
	\caption{Accumulated offsets provided by the state-of-the-art IDS and the NTP-based IDS with batch sizes of $20$ and $30$.
	The same portion of the data trace (with ID=$25$) from the Toyota dataset is used.
	Significant differences in slopes (i.e., estimated clock skew) are observed for the same message using the state-of-the-art IDS, whereas the clock skew estimated by the NTP-based IDS is almost identical with different batch sizes.
	}
	%
	%Fig. \ref{fig:counter_example:startingIndex} shows that slopes of messages 0x223 and 0x224 swap each other as we change the starting index of the input data from 1 to 2.
	%
	%Fig. \ref{fig:counter_example:differentN} shows that clock skew of message 0x020 changes from 273ppm to 151ppm as the number of messages in a batch is changed from 20 to 30.
	%
	%Comparing Fig. \ref{fig:toyota_accoffset_shin:run25} and Fig. \ref{fig:counter_example:run32}, the clock skew of message 0x020 in Run 25 is 273ppm whereas that in Run 32 is 306ppm.
	\label{fig:cho_shin_impact_of_batch_size}
\end{figure}

Fig.~\ref{fig:cho_shin_impact_of_batch_size} illustrates the accumulated offsets estimated by the two CDIS's with different batch sizes\footnote{Due to the lack of ground truth, the authors in \cite{Shin:2016:finger} empirically identified that 0x020, 0x0B2, 0x223 and 0x224 are transmitted by two different ECUs. However, based on our NTP-based clock skew estimation results, we believe that the four messages come from the same ECU.}. 
Significant differences in slopes for the same message are observed for the state-of-the-art IDS. 
For example, the estimated clock skew (based on the end point) of message 0x020 is around $273$ ppm with $N=20$, but dropped to around $151$ ppm with $N=30$.
In contrast, the NTP-based IDS  provides  consistent estimation. 

To further quantify estimation consistency, we consider the following three cases: 1) use the same portion of the same trace, and vary $N$ from $20$ to $100$ with a step of $20$, 2) set $N=20$, and use different portions of the same trace by omitting the first $m$ messages, where $m$ is varied from $1$ to $19$, and 3) set $N=20$, and use $14$ different traces from the Toyota dataset. 
The standard deviation ($\sigma$) of estimated clock skews are adopted as the metric, and a smaller $\sigma$ value implies more consistent estimation.
As shown in Table~\ref{table:cho_shin_impact_of_batch_size}, the NTP-based IDS has a significantly smaller $\sigma$ than the state-of-the-art IDS for all messages in all cases.

\begin{table}[ht!]
	\caption {Standard deviations ($\sigma_1$, $\sigma_2$, $\sigma_3$) of clock skews estimated by IDS in three different cases.
		The NTP-based IDS has a significantly smaller $\sigma$ than the state-of-the-art IDS, which demonstrates its consistency in clock skew estimation.
	}
	%as the batch size changes from 20 to 100 by step size of 20 using Toyota data. Unit is in ppm.}
	\label{table:cho_shin_impact_of_batch_size}
	\begin{center}
		\resizebox{1\columnwidth}{!}
		{
			\begin{tabular}{|c|c|c|c|c|c|c|}
				\hline
				\multirow{2}{*}{Message ID} & \multicolumn{3}{c|}{State-of-the-art IDS} &
				\multicolumn{3}{c|}{NTP-based IDS} \\
				\cline{2-7}
				& $\sigma_1$ & $\sigma_2$ & $\sigma_3$ & $\sigma_1$ & $\sigma_2$ & $\sigma_3$  \\
				\hline
				0x020 & 92.3682 & 12.0589 & 20.1727 & 0.3706 & 0.2000 & 1.7716 \\
				\hline
				0x0B2 & 94.4480 & 11.7543 & 19.4549 & 0.4252 & 0.2045 & 1.7929 \\
				\hline
				0x223 & 41.6631 & 16.2060 & 25.4885 & 0.3083 & 0.4429 & 1.6437 \\
				\hline
				0x224 & 29.0736 & 17.0442 & 32.6059 & 0.8348 & 0.5491 & 2.3660\\
				\hline
				0x2C1 & 85.3753 & 7.9963 & 26.2618 & 0.0866 & 1.2191 & 3.2977\\
				\hline
				0x2C4 & 116.5630 & 13.0896 & 53.7820 & 1.0763 & 1.1599 & 3.3602\\
				\hline
			\end{tabular}
		}
	\end{center}
\end{table}
\normalsize

\section{Proposed Cloaking Attack}
\label{sec:cloaking_attack}
%A IDS monitors inter-arrival time of a periodic message to track its clock skew and detect intrusion.
In this section, we propose the \textit{cloaking attack}, an intelligent masquerade attack, in which the adversary adjusts the inter-departure time of spoofed messages to manipulate the estimated clock skew as well as correlation to bypass an IDS.

\subsection{Cloaking Attack on Clock Skew Detector}
%Before explaining the cloaking attack, let us first understand why the masquerade attack can be easily detected by a IDS.
%As explained in Section \rev{XX}, the state-of-the-art IDS uses the average inter-arrival time of the previous batch to estimate the arrival time of each message in the current batch, and computes the average difference.
%If we ignore all uncertainties due to network delay and quantization errors etc., the metric in Eq.~(\ref{eq:cho_shin_avg_offset}) should be zero, because the messages are transmitted exactly every $T$ sec as per the transmitter's clock, and received every $\frac{T}{1+S}$ sec as per the receiver's clock, where $S$ is the transmitter's clock skew. 
Consider a message transmitted by the targeted ECU B every $T$ sec (e.g., $20$ms) in its own clock, which corresponds to  every $\hat{T}=T/(1+S_B)$ sec in the receiver R's clock, where $S_B$ is B's clock skew.
For the ease of discussion, we ignore offset deviations and the noise in arrival timestamps due to network delay and quantization.
Then B's clock skew as estimated by R is given by $\hat{S} = (T-\hat{T})/\hat{T} = S_B$.

In the masquerade attack, the weak attacker prevents the targeted message from being transmitted by ECU B. The strong attacker, which controls ECU A, transmits the false message every $T$ seconds, as measured by $C_{A}$ (Fig. \ref{fig:masquerade_attack_model}). Hence, ECU R receives the message every $\hat{T}^{\prime} = T/(1+S_{A})$ seconds, as measured by $C_{R}$, where $S_{A}$ is the clock skew between $C_{R}$ and $C_{A}$. The clock skew measured by ECU R for the messages injected by the attacker will then be $\hat{S}^{\prime} = S_{A}$. Therefore, if $S_{A} \neq S_{B}$, then the IDS will detect a change in the clock skew after the adversary begins transmitting.

The insight underlying our attack is that while the clock skew is a physical property, clock skew estimation in any IDS is based entirely on message inter-arrival time, which can be easily manipulated by the transmitter (i.e., the strong attacker controlling ECU A) adjusting the message inter-departure time. Effectively, the adversary \emph{cloaks} the skew of its hardware clock, % by using the software clock, 
thus motivating the term \emph{cloaking attack}. Under the cloaking attack, instead of transmitting every $T$ seconds, the attacker-controlled ECU A transmits every $\tilde{T}=T + \Delta T$ seconds, in order to match the clock skew observed at R.

%In a masquerade attack, B's is prevented from transmitting the target message, and the impersonating ECU A start transmitting the message every $T$ sec (i.e., $20$ms) in its own clock, which corresponds to $\hat{T}'=\frac{T}{1+S_A}$ in R's clock, where $S_A$ is A's clock skew.  
%Then R's estimate for the transmitter's clock skew will be $\hat{S}'=S_A$. 
%Therefore, if A and B have different skews, i.e., $S_A \neq S_B$, the inter-arrival time and the estimated clock skews will be different, i.e., $\hat{T}' \neq \hat{T}$ and $\hat{S}' \neq \hat{S}$, which is very likely to cause the IDS to raise an alarm.
%In practice, the detection capability of a IDS also depends on the degree of discrepancy between A and B's clock skews, offset variations, as well as noise. 

%Although clock skew is a physical property that cannot be changed, clock skew estimation at a IDS is purely based on message inter-arrival time, which can be easily manipulated by the transmitter by adjusting message inter-departure time. 
%That is, by adjusting the software clock for message periodicity, the adversary is able to cloak the skew of the hardware clock, and bypass the IDS.
%Hence, it is referred to as the \textit{cloaking attack}.
%It turns out that the cloaking strategy is rather simple: 
%\textit{instead of every $T$ sec, it transmits the spoofed message every $\tilde{T}=T+\Delta T$ sec as per the impersonating ECU's local clock, where $\Delta T \ll T$ is the amount of time adjusted}.

%With this strategy, the inter-arrival time seen by R would be 
The choice of $\Delta T$ is discussed as follows. Under the cloaking attack, the inter-arrival time observed by R is 
\begin{equation*}
\hat{T}'' = \frac{\tilde{T}}{1 + S_A} = \frac{T+\Delta T}{1 + S_A}
\end{equation*}
and the transmitter's clock skew estimated by R is
\begin{equation}
\hat{S}'' = \frac{T - \hat{T}''}{\hat{T}''} = \frac{S_A \cdot T - \Delta T}{T+\Delta T}.
\end{equation}
Hence, to bypass the IDS, the adversary needs to choose  $\Delta T$ such that $\hat{S}''=\hat{S}$, or equivalently $\hat{T}'' = \hat{T}$, which means
\begin{equation}
\Delta T = \frac{(S_A - S_B)}{1+S_B} \cdot T = S_{AB} \cdot T = \frac{-S_{BA}}{1+S_{BA}}\cdot T, \label{eq:delta_T}
\end{equation}
where $S_{AB}$ is A's clock skew relative to B's clock, and the last two equalities are due to Eq.~(\ref{eq:relative_skew}) and Eq.~(\ref{eq:relative_skew_conversion}), respectively.

Therefore, the message inter-departure time $\tilde{T}$ would be
\begin{equation*}
	\tilde{T} = T + \Delta T = T - \frac{S_{BA}}{1+S_{BA}} T = \frac{T}{1+S_{BA}},
\end{equation*}
which is the period of the message from B (i.e., weak attacker) measured by the local clock of A (i.e., strong attacker). 

To summarize, the cloaking attack is performed as follows. 
%\begin{itemize}
	%\item
	 After the adversary compromises two ECUs as strong and weaker attackers, the strong attacker estimates the period of the target message $\tilde{T}$ as measured by its local clock. 
	%\item 
	During the masquerade attack, the strong attacker transmits the spoofed message every $\tilde{T}$ sec. 
%\end{itemize}
While the preceding analysis ignores the noise present in the system, our results in Section \ref{sec:evaluation} show that the cloaking attack is effective in a realistic environment.
%In practice, however, noise unavoidably affects the estimation accuracy of $\tilde{T}$ at the strong attacker, as well as the clock skew estimation accuracy at the receiver. 
%But as we will show in Section \rev{XX}, the proposed cloaking attack will significantly increase the chance of bypassing the detection of a IDS. 

\subsection{Maximum Slackness Index (MSI)}
\label{subsec:MSI}
%The goal of the simulation was to determine the range of $\Delta T$ values for which the cloaking attack was effective against each IDS. 
In practice, the adversary will be unable to precisely match the clock skew of the targeted ECU due to hardware limitations. Deviations between the clock skew of the attacker and the targeted ECU, however, may still be mistaken for random delays and quantization errors by the IDS. These sources of randomness create an interval of $\Delta T$ that an adversary can introduce while remaining undetected; the more effective the detector, the smaller the interval of $\Delta T$ will be. We introduce a metric that formalizes this notion  as follows. 
%The inherent randomness of the in-vehicle network, however, means that deviations between the clock skew of the attacker and the targeted ECU may be mistaken for random delays and quantization errors. 
%In order to quantify this effectiveness, 
We first let $P_{s}(\Delta T)$ 
%$P_{s}(\mathcal{D},\Delta T)$ 
denote the probability of a successful cloaking attack 
%against detector \hl{$\mathcal{D}$} 
when the added delay is $\Delta T$. We define the upper and lower limits of $\Delta T$ for a successful attack as 
\begin{align*}
(\Delta T)_{\max}(\epsilon)&=\max{\{\Delta T: P_{s}(\Delta T) > 1-\epsilon\}} \\
(\Delta T)_{\min}(\epsilon)&=\min{\{\Delta T: P_{s}(\Delta T) > 1-\epsilon\}}
\end{align*}

We define the $\epsilon$-\textit{Maximum Slackness Index} ($\epsilon$-MSI) of the attacker %against detector $\mathcal{D}$ 
as 
%\begin{equation*}
$\epsilon\text{-MSI} = (\Delta T)_{max}(\epsilon) - (\Delta T)_{min}(\epsilon).$ 
%\epsilon-MSI(\mathcal{D}) = \max{\{|\hat{\Delta T}| : \mbox{ probability of success }}\\ \mbox{against detector $\mathcal{D}$ with offset } \overline{\Delta T} + \hat{\Delta T} > (1-\epsilon)\},
%\end{equation*}
%where $\overline{\Delta T}$ is the value of $\Delta T$ obtained in Eq. (\ref{eq:delta_T}). In words $\epsilon\text{-MSI}$ is the largest deviation from the ideal clock skew $\overline{\Delta T}$ that still results in an attack success probability exceeding $(1-\epsilon)$. 
 The normalized $\epsilon\text{-MSI}$ is defined as the ratio between of $\epsilon\text{-MSI}$ (in $\mu$s) and the message period (in sec), and its unit is ppm.
Intuitively, a smaller value of $\epsilon\text{-MSI}$ signifies a more effective detector and less freedom for the attacker, since the adversary's clock skew must closely match the targeted ECU's in order to remain undetected.

\subsection{Cloaking Attack on Correlation Detector}\label{sec:cloaking_on_correlation_analysis}
If the spoofed message has a sibling message with highly correlated offsets, the correlation detector can be deployed as the secondary countermeasure. 
Before introducing the cloaking attack on the correlation detector, let us discuss why two messages consecutively transmitted and consecutively received are more likely to have high correlation in average offsets.
Due to space constraints, we focus on the NTP-based IDS, but the same logic is applicable to the state-of-the-art IDS.

%Since the two messages are consecutively transmitted and received, it is reasonable to assume that each pair 

Denote the $i$-th message in the $k$-th batch for messages $v$ and $w$ as $v_{k,i}$ and $w_{k,i}$, which are transmitted at $t_{k,i}^{(v)}$ and $t_{k,i}^{(w)}$, respectively.\footnote{This is another requirement for two messages to be highly correlated: the two consecutively transmitted messages needs to be processed as simultaneously as the $i$-th message in the $k$-th batch.}
Without loss of generality, suppose that $w_{k,i}$ is transmitted right after $v_{k,i}$.
Let $\Delta t$ be the transmission duration of each message $v$, which is constant, given the fixed message length and CAN bus speed. 
Hence, we have $t_{k,i}^{(w)}=t_{k,i}^{(v)}+\Delta t$.

Let us consider the first case where $v_{k,i}$ and $w_{k,i}$ are received consecutively at $a_{k,i}^{(v)}$ and $a_{k,i}^{(w)}$, which means no other messages with higher priority IDs are received between $a_{k,i}^{(v)}$ and $a_{k,i}^{(w)}$ due to arbitration. 
For simplicity, we assume constant network delays for both messages (denoted as $d_v$ and $d_w$, respectively), and ignore quantization noise at the receiver. 
Therefore we have $a_{k,i}^{(w)}=a_{k,i}^{(v)}+\Delta t + (d_w - d_v)$.

In the NTP-based IDS, the estimated average offset for messages $v$ and $w$ in the $k$-th batch are  
\begin{eqnarray}
\nonumber
O_{avg}^{(v)}[k] &=& T-\frac{1}{N}\left( a_{k,N}^{(v)} - a_{k,0}^{(v)} \right) \\
\nonumber
 &=& -O^{(v)} - \frac{1}{N}\left(\epsilon_{k,N}^{(v)}-\epsilon_{k,0}^{(v)}\right) \\
O_{avg}^{(w)}[k] &=& T-\frac{1}{N}\left( a_{k,N}^{(w)} - a_{k,0}^{(w)} \right) = O_{avg}^{(v)}[k]. \label{eq:cloaking_correlation_same_avg_offset}
\end{eqnarray}
Since $O_{avg}^{(v)}[k]$ and $O_{avg}^{(w)}[k]$ are the $k$-th realizations of the random variables $O_{avg}^{(v)}$ and $O_{avg}^{(w)}$, respectively, Eq.~(\ref{eq:cloaking_correlation_same_avg_offset}) implies $O_{avg}^{(w)}=O_{avg}^{(v)}$, and thus their correlation coefficient $\rho$ is as high as $1$.
In general, as along as the two messages are received with a constant delay (consecutive reception is a special case), they will have high correlation. 
In practice, however, the correlation would slightly decrease due to network delay variations and quantization noise at the receiver.

Next we examine the second case in which messages with higher priority IDs are received in between the two messages.
Let the arbitration delay be $d_{k,i}\geq 0$, and thus $a_{k,i}^{(w)}=a_{k,i}^{(v)}+\Delta t + (d_w - d_v) + d_{k,i}$.
Then we have
\begin{equation}
O_{avg}^{(w)}[k] = O_{avg}^{(v)}[k] - \frac{1}{N}(d_{k,N}-d_{k,0}),
\end{equation}
where the second term may be considered as the $k$-th realization of a random variable $D$, independent of $O_{avg}^{(v)}$ and $O_{avg}^{(w)}$. 
Therefore, we have $O_{avg}^{(w)}=O_{avg}^{(v)} + D$, and 
\begin{equation*}
\rho \left(O_{avg}^{(v)}, O_{avg}^{(w)} \right) = \frac{\sqrt{Var (O_{avg}^{(v)})}}{\sqrt{Var(O_{avg}^{(v)}) + Var(D)}} < 1.
\end{equation*}
As a result, depending on the variance of arbitration delay, the correlation in the second case may be much smaller than $1$.

On the other hand, if two messages are transmitted from different ECUs, we have $O_{avg}^{(w)}[k]=-O^{(w)}- \frac{1}{N}\left(\epsilon_{k,N}^{(w)}-\epsilon_{k,0}^{(w)}\right)$. 
Since $\{\epsilon_{k,i}^{(v)}\}$ and $\{\epsilon_{k,i}^{(w)}\}$ are independent, $O_{avg}^{(w)}$ is also independent of $O_{avg}^{(v)}$, which implies $\rho\approx 0$.
The above analysis is supported by our hardware evaluation (Section \ref{sec:evaluation_attack_on_correlation}).

Hence, in order to thwart the correlation detector, the attacker adopts the following strategy. Before executing the attack, the attacker observes the targeted message over a period of time and identifies any sibling messages. 
To launch the cloaking attack, the strong attacker-controlled ECU A begins transmitting the targeted message immediately after the sibling message is completed. Since the transmission from ECU A begins once the sibling message transmission ends, the average offset of the targeted and sibling messages will be equivalent and show high correlation, as described by Eq. (\ref{eq:cloaking_correlation_same_avg_offset}).
Note that Eq. (\ref{eq:cloaking_correlation_same_avg_offset}) also implies that their accumulated offsets as well as estimated clock skews will be equivalent, thus bypassing the clock skew detector at the same time.

%To bypass the correlation detector, the attack strategy is also rather simple: \textit{the impersonating ECU transmits the spoofed message immediately after it observes a sibling message of the spoofed message.}
%\cmt{Explain that this is to make sure that the two messages are consecutively received by the IDS with a constant delay.}
%According to our previous discussion, this strategy should lead to high correlation between the two messages.
%In fact, \rev{it also ensures that they have very close clock skews (Eq.~(\ref{eq:cloaking_correlation_same_avg_offset})), and can bypass the clock skew detector.}

%>>> SECTION: Evaluation
%\input{./sections/evaluation}
\section{Evaluation}
\label{sec:evaluation}
In this section, we  evaluate the performance of the proposed cloaking attack on two CAN bus testbeds, and demonstrate that the cloaking attack is able to bypass both the state-of-the-art and the NTP-based IDSs.
We first describe our testbeds, followed by an illustration of a single trial run of our proposed attack. We then give detailed results for the cloaking attack against both the clock skew and correlation detectors.

\subsection{Testbeds}
We built two CAN bus testbeds: a CAN bus prototype and a CAN testbed on a real vehicle (University of Washington (UW) EcoCar\footnote{The EcoCar was originally a 2016 Cherolet Camaro donated by General Motors as part of a competition.
	Researchers at UW later converted into a hybrid electric vehicle to develop next-generation vehicle technologies. } \cite{ecocar}).
Compared with the prototype with three ECUs, the EcoCar testbed hosts $18$ stock ECUs and one ECU added by researchers.
There are a total of $89$ messages with different IDs, and at least $2500$ messages are transmitted every second. 
The EcoCar testbed provides a real CAN environment to evaluate and demonstrate the proposed cloaking attack. 

\subsubsection{CAN Bus Prototype}

% Testbed and data collection
As shown in Fig.~\ref{fig:arduino_testbed}, our CAN bus prototype consists of three ECUs.
Each ECU is composed of an Arduino UNO board and a Sparkfun CAN bus shield.
%
%Arduino UNO board creates the contents of a CAN message and determines the period of a message.
%
%Sparkfun CAN bus shield constructs a CAN message using data from Arduino UNO board, and decodes a CAN message.
%
The CAN bus shield uses a Microchip MCP2515 CAN controller, a Microchip MCP2551 CAN transceiver, and a 120$\Omega$ terminator resistor. 
The bus speed of the prototype is set to $500$Kbps as in typical  CAN buses.
ECU $1$ is the receiving ECU that implements the IDS.
ECU $2$ is the targeted ECU controlled by the weak attacker that transmits messages 0x11 every $100$ms (i.e., $10$Hz).
ECU $3$ is the strong attacker that aims to impersonate ECU $2$ in a masquerade or cloaking attack.

\subsubsection{EcoCar CAN testbed}
\begin{figure}[t!]
	\centering
	\begin{subfigure}[h]{0.43\columnwidth} % {0.48\columnwidth}
		\includegraphics[width=\columnwidth]{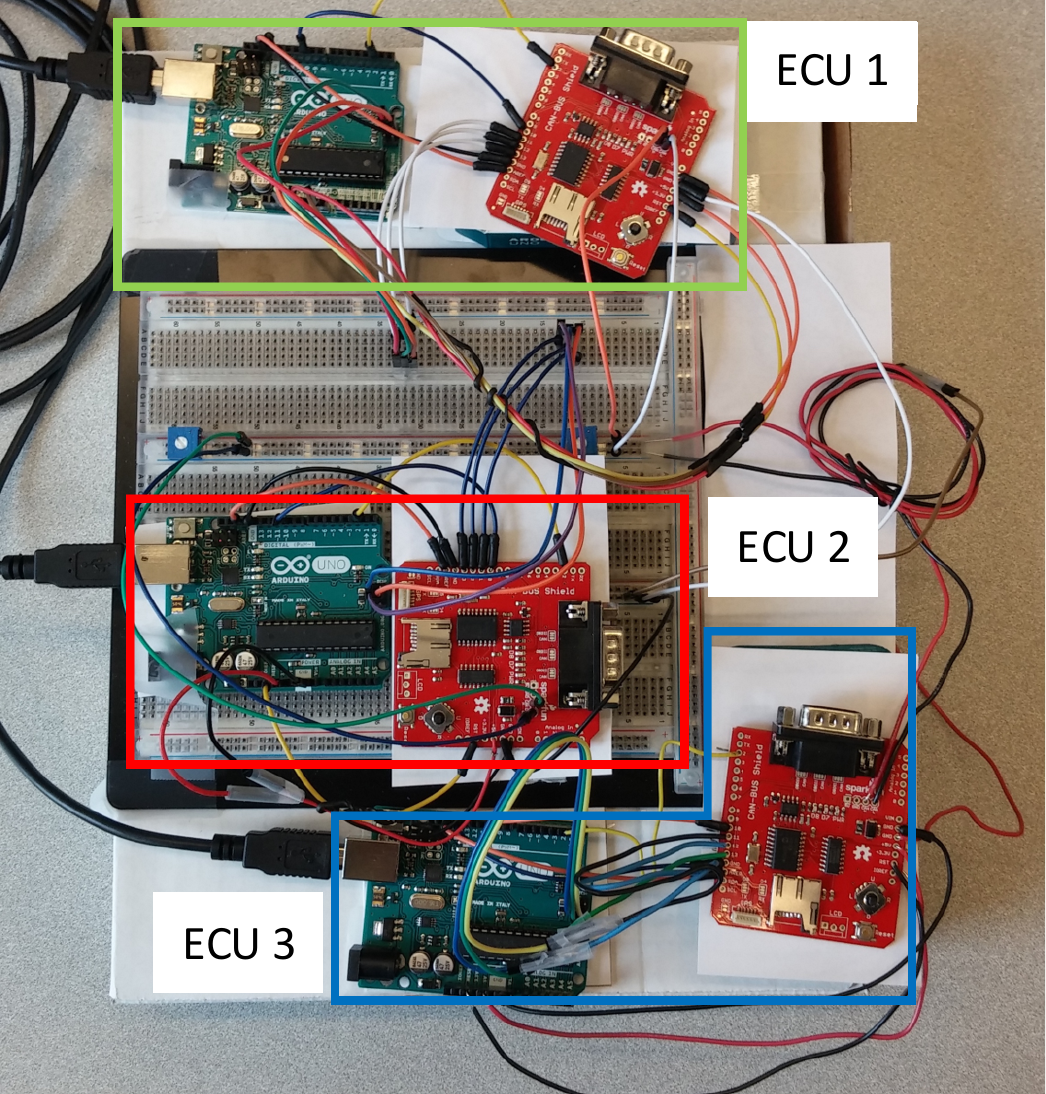}
		\caption{CAN bus prototype}
		\label{fig:arduino_testbed}
	\end{subfigure}
	\begin{subfigure}[h]{0.55\columnwidth} % {0.48\columnwidth}
		\includegraphics[width=\columnwidth]{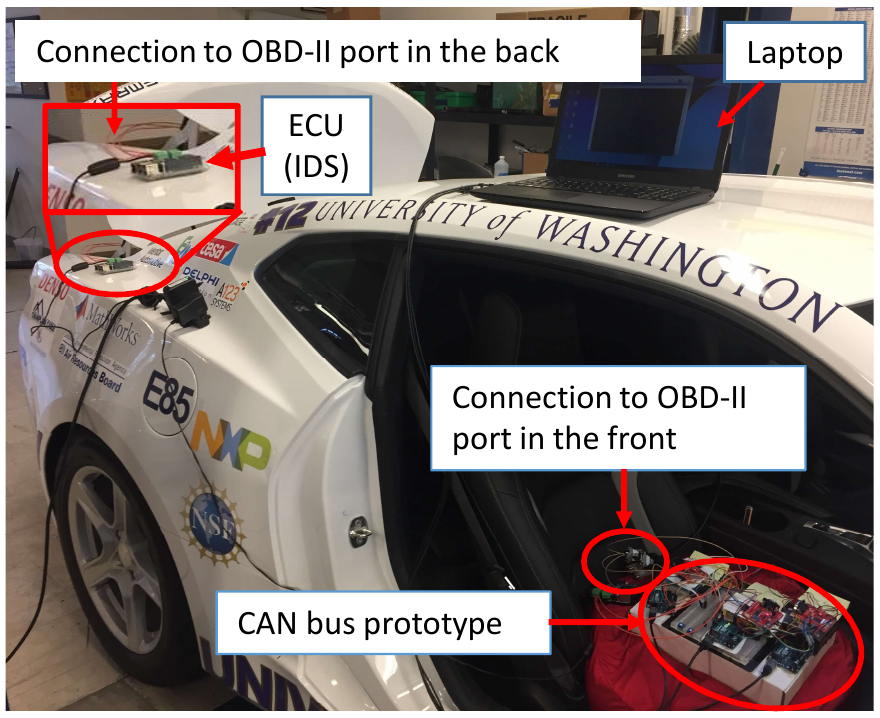}
		\caption{EcoCar testbed}
		\label{fig:real_vehicle_testbed}
	\end{subfigure}
	\caption{CAN bus testbeds. The CAN bus prototype is connected to the CAN bus inside the EcoCar via the OBD-II port to build the EcoCar testbed.  }
\end{figure}

As shown in Fig. \ref{fig:real_vehicle_testbed}, the CAN bus prototype is connected to the in-vehicle CAN bus of the EcoCar via the On-Board Diagnostics (OBD-II)  port to build the EcoCar testbed. 
During our experiments, the EcoCar is in park mode in an isolated and controlled environment, but all ECUs are functional and actively exchange CAN messages.

Due to the large CAN traffic and limited computing capability, the Arduino-based ECU is not able to capture all messages. 
Hence, we build a fourth ECU that consists of a Raspberry Pi 3 and a PiCAN 2 board (which has the same CAN controller and transceiver as in the CAN bus shield) as the receiving ECU. 
A stock ECU is considered as the targeted ECU (the weak attacker) which transmits message 0x184 every $100$ ms (i.e., $10$Hz), and the same Arduino-based ECU $3$ is used as the strong attacker that injects spoofed messages.

\subsection{Example of NTP-based IDS}\label{sec:evaluation_example}
For illustration, we first describe a single execution of the masquerade attack and the behavior of the NTP-based IDS. We compare the masquerade attack without cloaking and our proposed cloaking attack.  
In the example, we set the update threshold $\gamma$ to $4$ and the detection threshold $\Gamma$ to $5$ for the NTP-based IDS. 
For data collected from the CAN bus prototype, the sensitivity parameter $\kappa$ is set to $5$. 

%As shown in Fig. \ref{fig:masquerade_attack_model}, t
The IDS first tracks the clock skew of message 0x11 from the targeted ECU for $1000$ seconds, before the attack happens. 
Then the IDS is fed with the timestamps of attack messages.
For the masquerade attack, the strong attacker transmits every $T=100$ ms according to its local clock. 
For the cloaking attack, the strong attacker first observes the inter-arrival time of message 0x11 to be around 100040 $\mu s$, and then adjusts the message inter-departure time to be $\tilde{T}=100040$ $\mu$s, where $\Delta T = 40$ $\mu$s.

\begin{figure}[t!]
	\centering
	\begin{subfigure}[h]{0.49\columnwidth} % {0.48\columnwidth}
		\captionsetup{justification=centering}
		\includegraphics[width=\columnwidth]{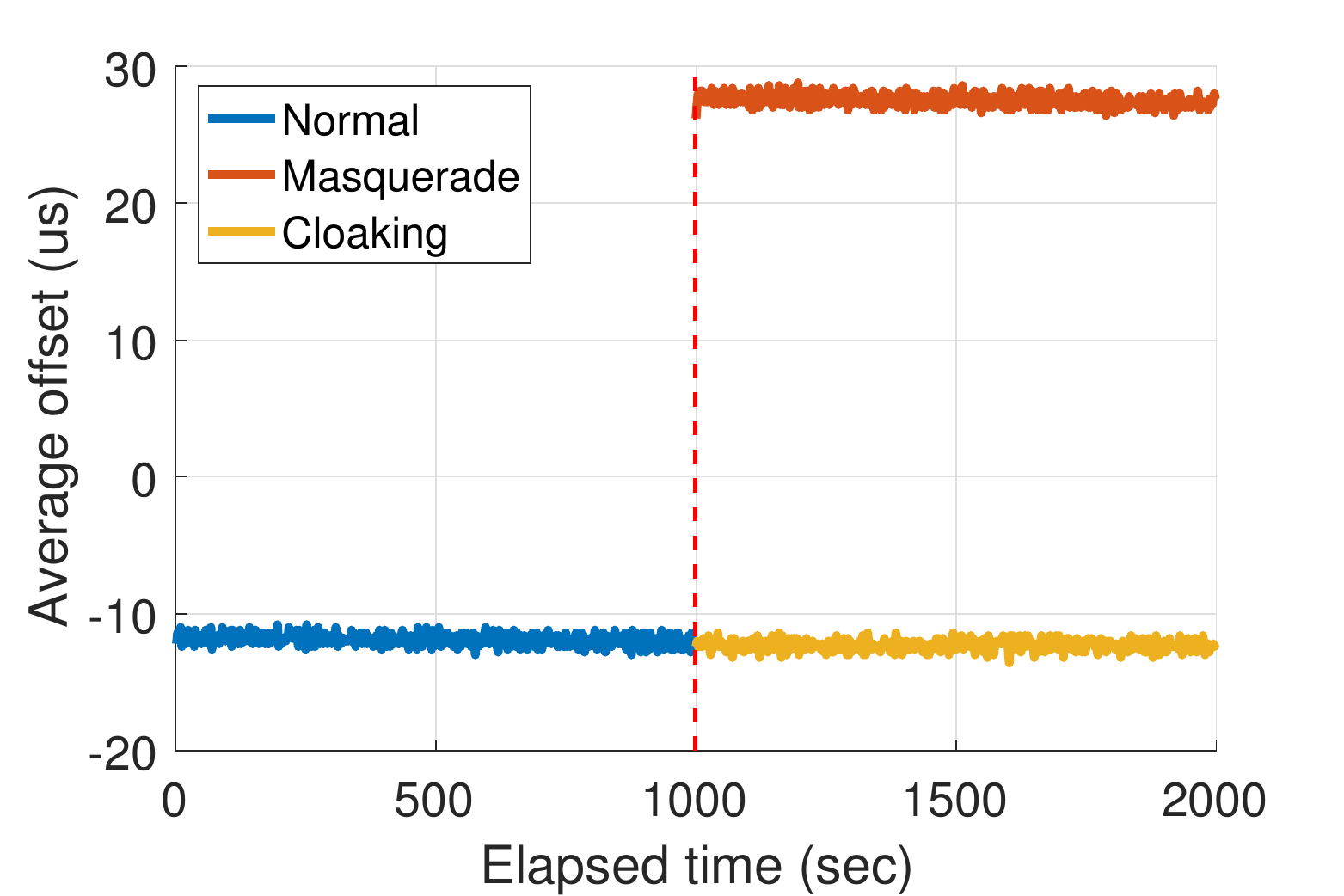}
		\caption{Average offset}
		\label{fig:arduino_avg_offset}
	\end{subfigure}
	\begin{subfigure}[h]{0.49\columnwidth} % {0.48\columnwidth}
		\captionsetup{justification=centering}
		\includegraphics[width=\columnwidth]{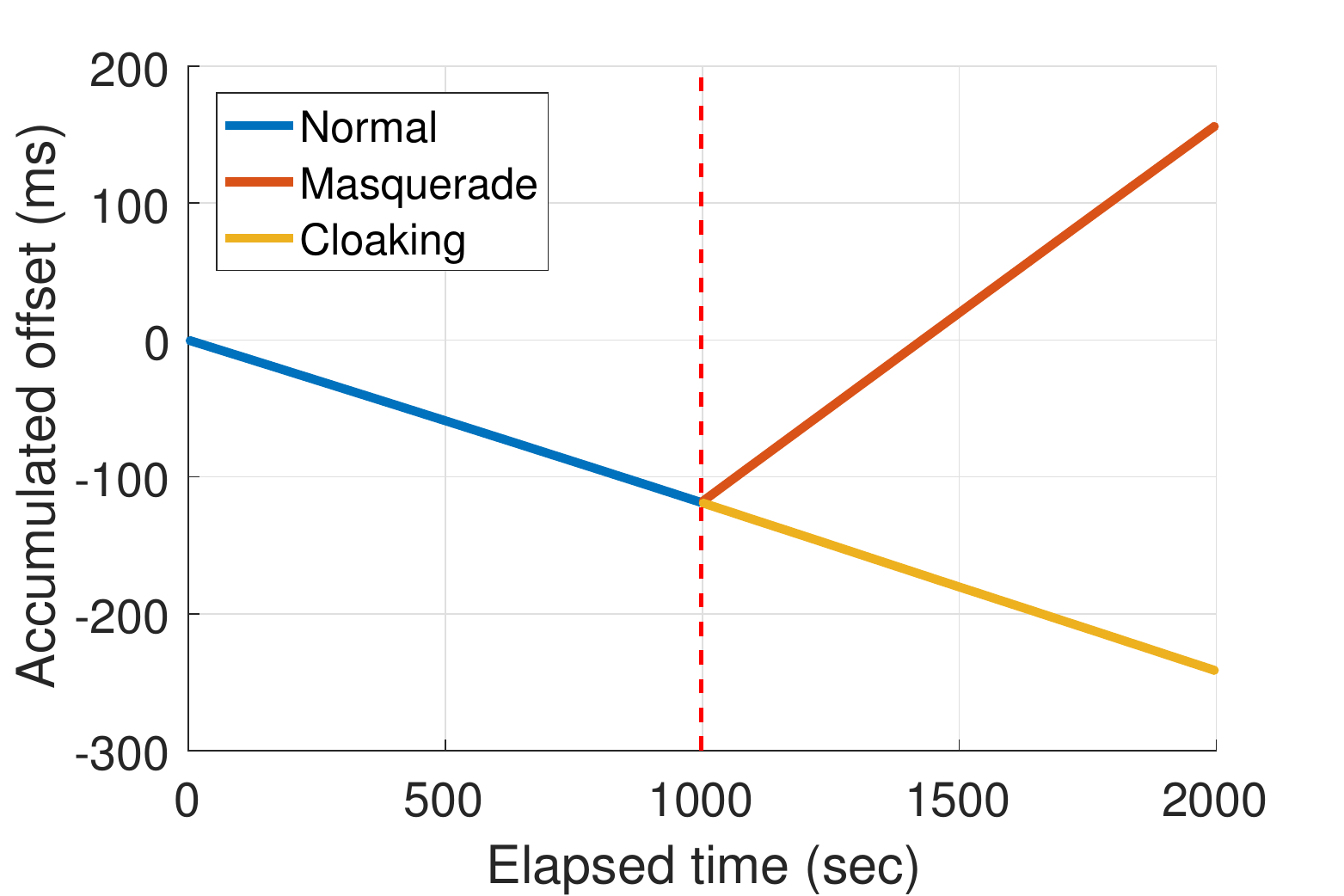}
		\caption{Accumulated offset}
		\label{fig:arduino_acc_offset}
	\end{subfigure}
	\\
	\begin{subfigure}[h]{0.49\columnwidth} % {0.48\columnwidth}
		\captionsetup{justification=centering}
		\includegraphics[width=\columnwidth]{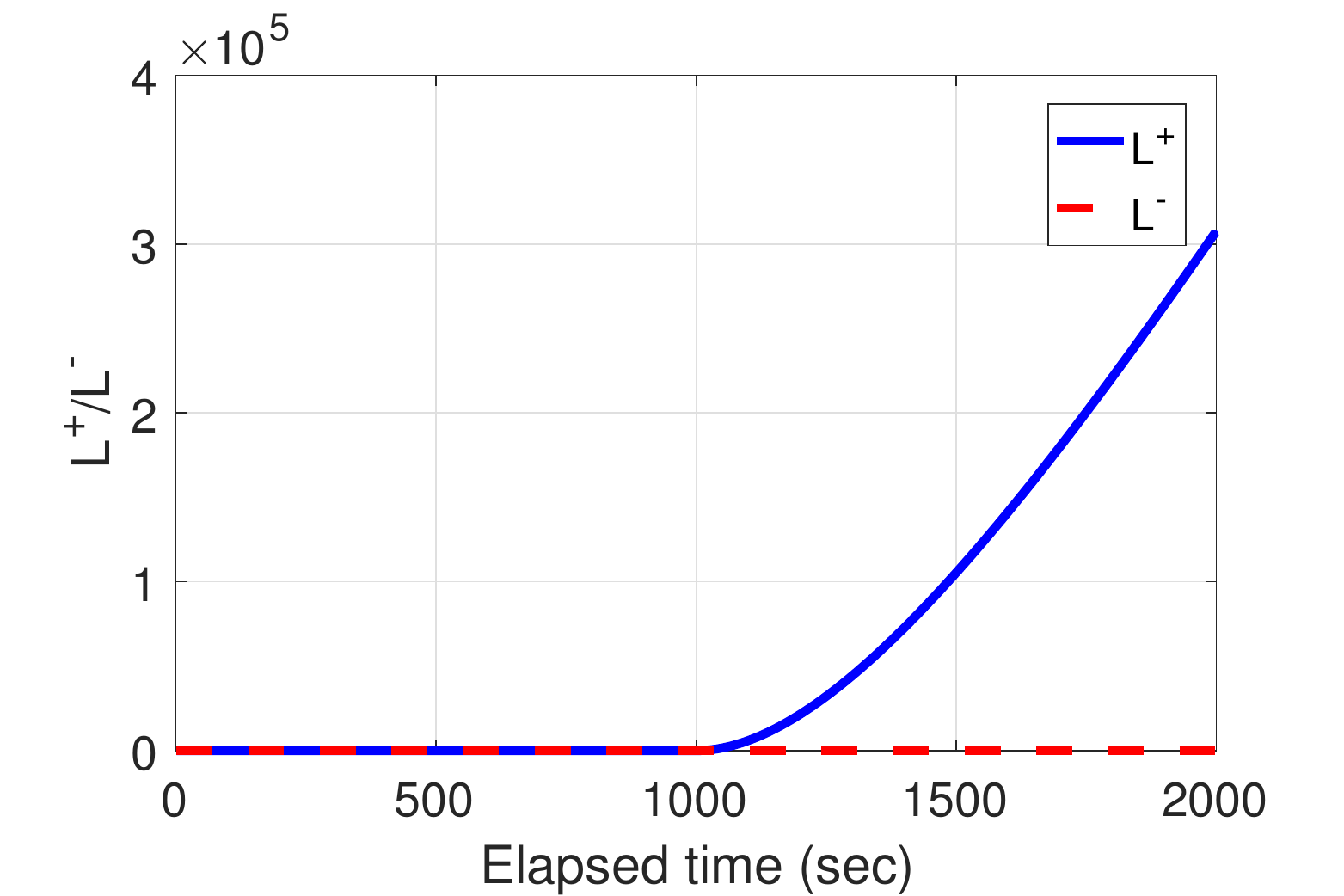}
		\caption{Control limits under the masquerade attack}
		\label{fig:arduino_control_limit_masquerade}
	\end{subfigure}
	\begin{subfigure}[h]{0.49\columnwidth} % {0.48\columnwidth}
		\captionsetup{justification=centering}
		\includegraphics[width=\columnwidth]{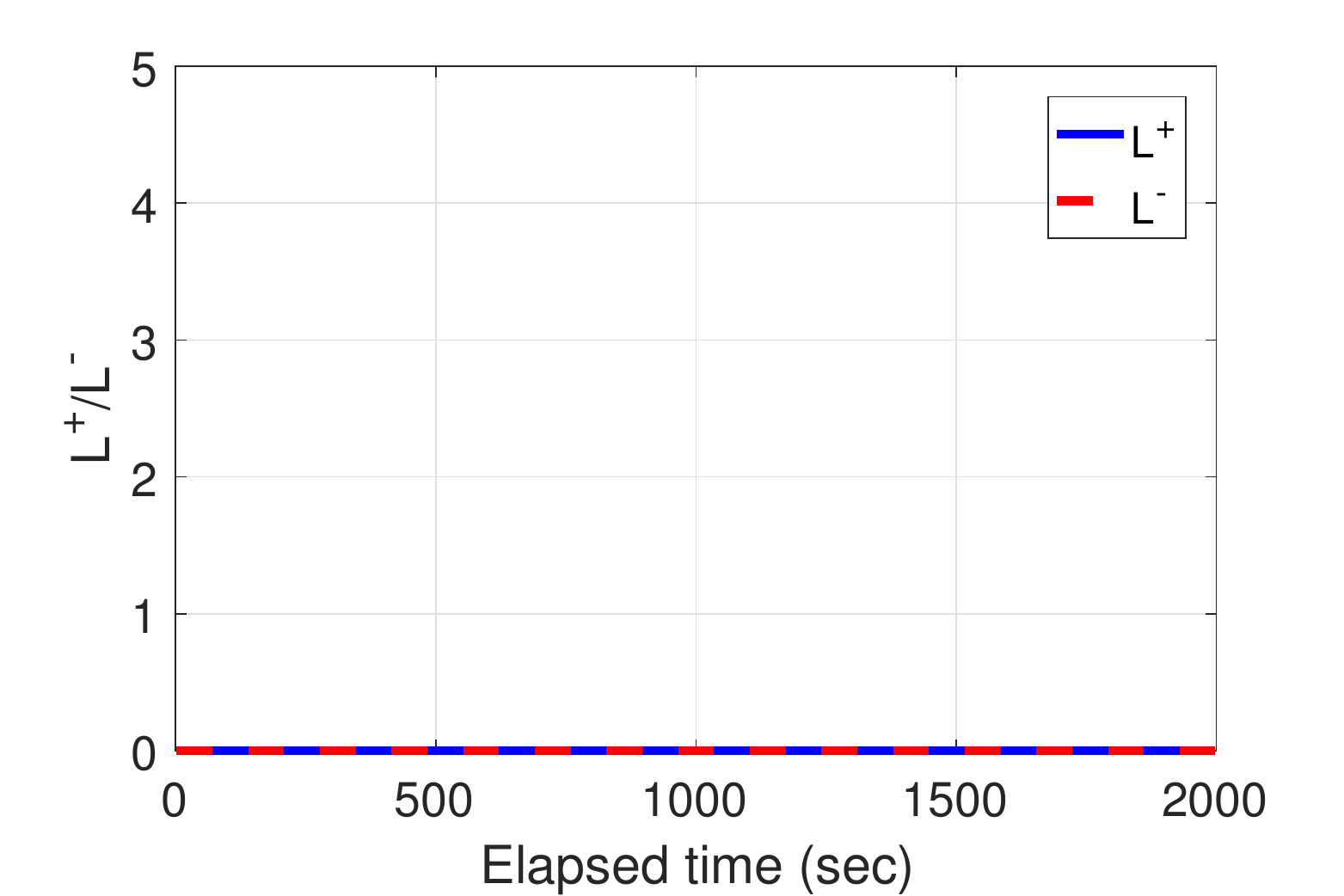}
		\caption{Control limits under the cloaking attack}
		\label{fig:arduino_control_limit_cloaking}
	\end{subfigure}
	\caption{Behavior of the NTP-based IDS under the masquerade and cloaking attacks on the CAN bus prototype, in terms of average offset, accumulated offset and control limits. In the masquerade attack, the accumulated offset grows over time and is detected by both IDS. Under the cloaking attack, the clock skews before and after the attack are indistinguishable.}
	\label{fig:arduino_one_experiment_example}
\end{figure}

As shown in Fig. \ref{fig:arduino_one_experiment_example}, when the masquerade attack happens, the average offset immediately jumps from around $-12 \ \mu$s to around $28 \ \mu$s (Fig. \ref{fig:arduino_avg_offset}), and the slope changes from $-118.9$ ppm to $275.3$ ppm (Fig. \ref{fig:arduino_acc_offset}), because of the very distinct clock skews between targeted and masquerading ECUs. 
As a result, such deviations add up and cause the control limits of the IDS to increase (Fig.~\ref{fig:arduino_control_limit_masquerade}).
In contrast, under the cloaking attack, the average offset stays almost the same as the original curve, as does the  slope of 
the accumulated offset. 
Since the deviations are so small, the control limits are always zero, and thus the IDS is unable to detect the cloaking attack (Fig.~\ref{fig:arduino_control_limit_cloaking}). Tests on the EcoCar testbed lead to similar observations.
%We do not provide an example from the EcoCar tested, since it leads to very similar observations. 
%Details about the behavior of the state-of-the-art IDS are available in \cite{Shin:2016:finger}.

%On the EcoCar CAN testbed, the strong attacker injects the message 0x180 with the added delay of -28$\mu s$ in order to attack the message 0x184.

\subsection{Performance of Cloaking Attack on Clock Skew Detector}\label{sec:evaluation_attack_on_clock_skew}

When launching the cloaking attack, the impersonating ECU (Arduino-based) transmits every $100040\mu$s ($\Delta T=40\mu$s) on the CAN bus prototype to spoof the $10$Hz message 0x11, and every $99971\mu$s ($\Delta T=-29\mu$s) to spoof the $10$Hz message 0x184 on the EcoCar testbed.
We collected a total of 3.7 hours and 8.5 hours of attack data from the CAN bus prototype and the EcoCar testbed, respectively.
%Although the dataset is collected for the cloaking attack with a particular $\Delta T$ value, an extra amount of delays is added to the message inter-arrival time so as to generate datasets for cloaking attacks with different $\Delta T$ values, which may be treated as general masquerade attacks. 

To simulate the cloaking attack, the IDS is fed with $1000$ batches of normal data, followed by $n_{attack}$ batches of attack data in each experiment. 
We assume perfect timing for the cloaking attack, i.e., the first attack message is received at the next expected time instant of the targeted message.
The impact of mistiming on the cloaking attack is studied in Appendix~\ref{appendix:mistiming}. 
An attack is successful if it is undetected by the IDS, and failed otherwise. 
A total of $100$ non-overlapping segments of size $n_{attack}$ are prepared from the attack data to simulate $100$ independent attacks. 
To measure the attack performance, we compute \textit{successful attack probability}, denoted as $P_s$, which is the percentage of experiments where the attack is successful.

We consider the state-of-the-art IDS and the NTP-based IDS with batch size equal to $20$.
For the state-of-the-art IDS, the update threshold $\gamma$ is set to $3$ and the detection threshold $\Gamma$ is $5$ \cite{Shin:2016:finger}.
For the NTP-based IDS, we use $\gamma=4$ and $\Gamma=5$. 
For the data collected from the CAN bus prototype, the sensitivity parameter $\kappa$ is set to $5$ for both IDSs.
It is set to $8$ for the data collected from the EcoCar testbed to avoid false alarms. 

% Successful Rates
\begin{figure}[t!]
	\centering
	\begin{subfigure}[h]{0.49\columnwidth} % {0.48\columnwidth}
		\captionsetup{justification=centering}
		\includegraphics[width=\columnwidth]{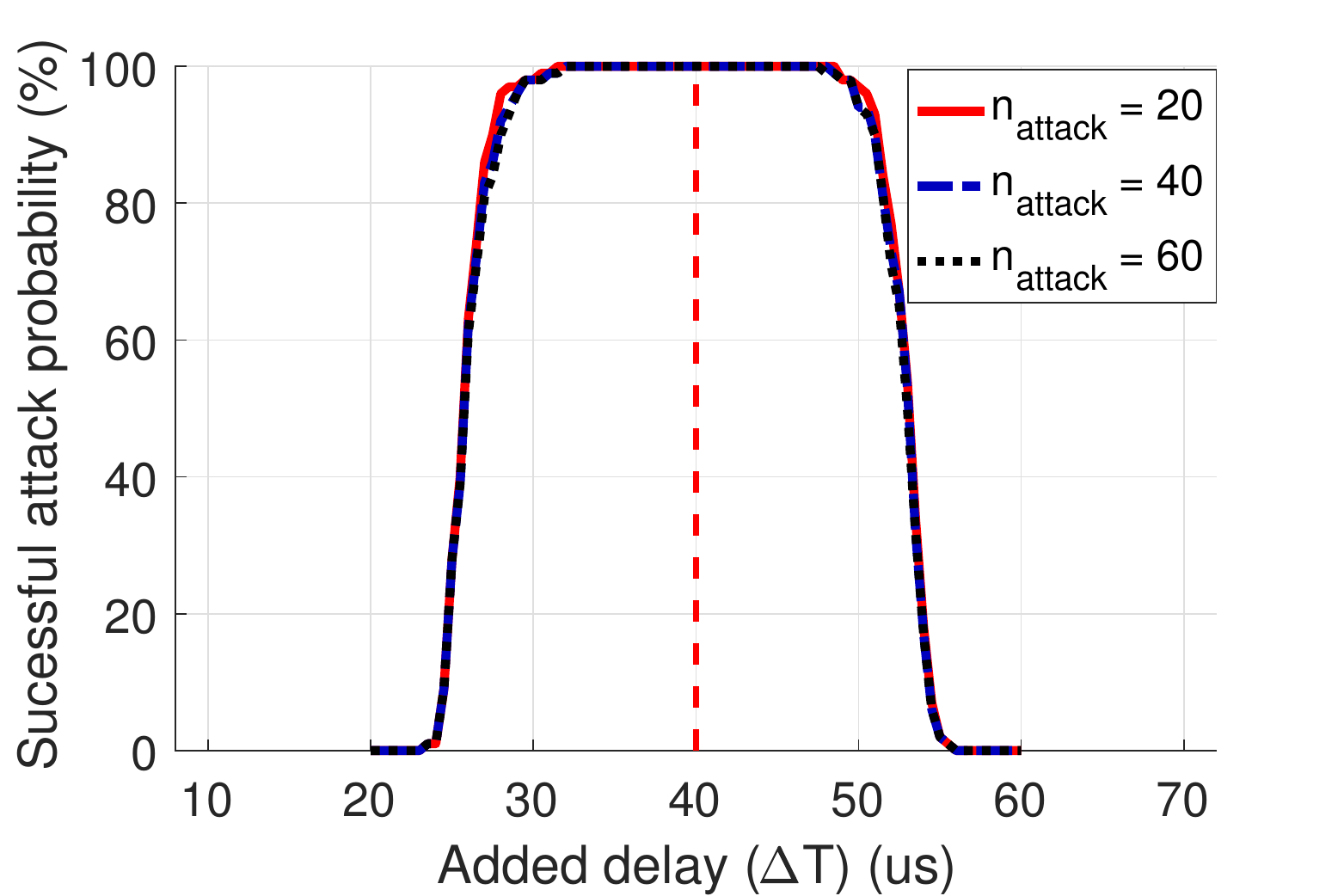}
		\caption{CAN prototype, state-of-the-art}
		\label{fig:arduino_clock_skew_attack_success_rate_Cho}
	\end{subfigure}
	\begin{subfigure}[h]{0.49\columnwidth} % {0.48\columnwidth}
		\captionsetup{justification=centering}
		\includegraphics[width=\columnwidth]{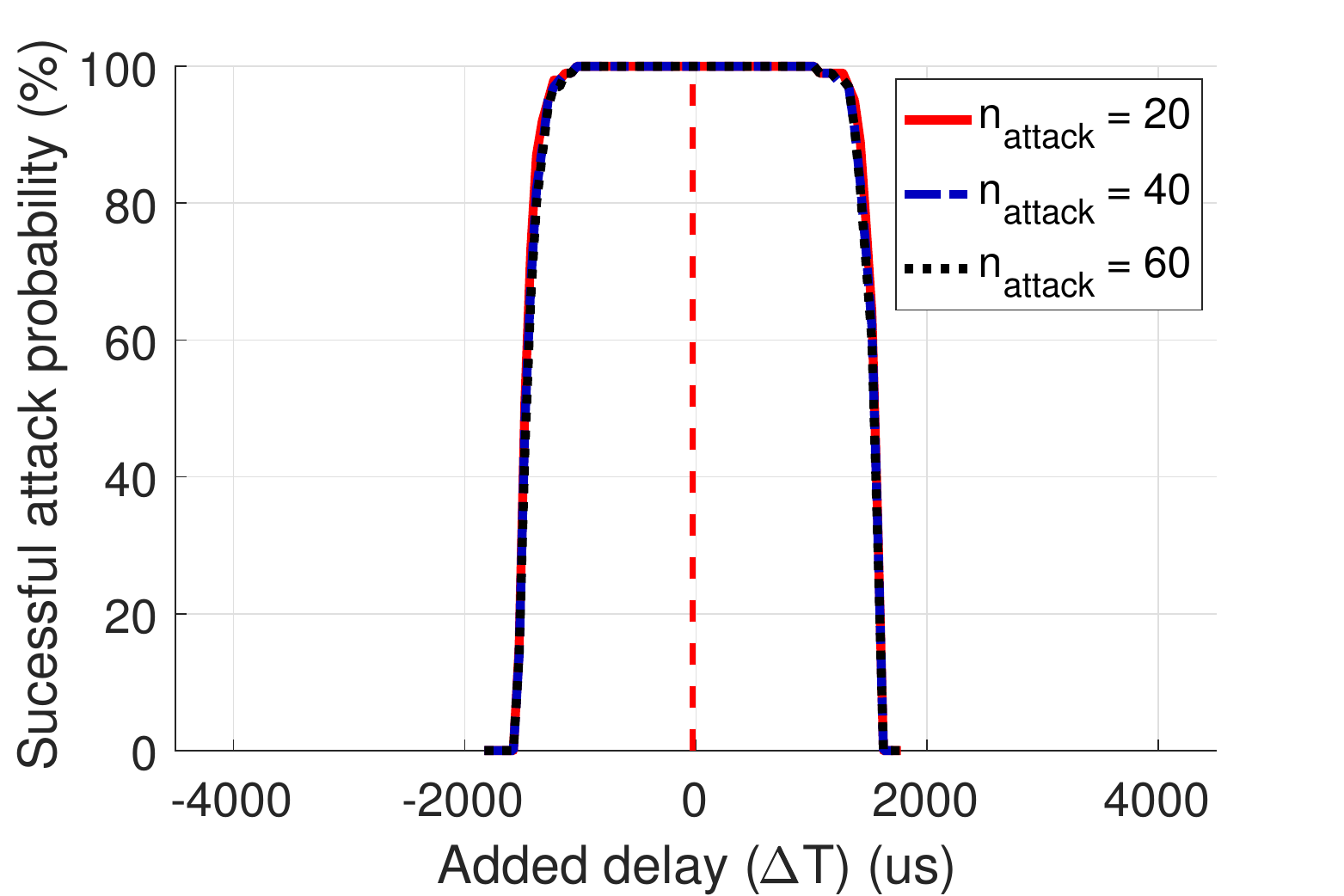}
		\caption{EcoCar testbed, state-of-the-art}
		\label{fig:ecocar_clock_skew_attack_success_rate_Cho}
	\end{subfigure}
	\\
	\begin{subfigure}[h]{0.49\columnwidth} % {0.48\columnwidth}
		\captionsetup{justification=centering}
		\includegraphics[width=\columnwidth]{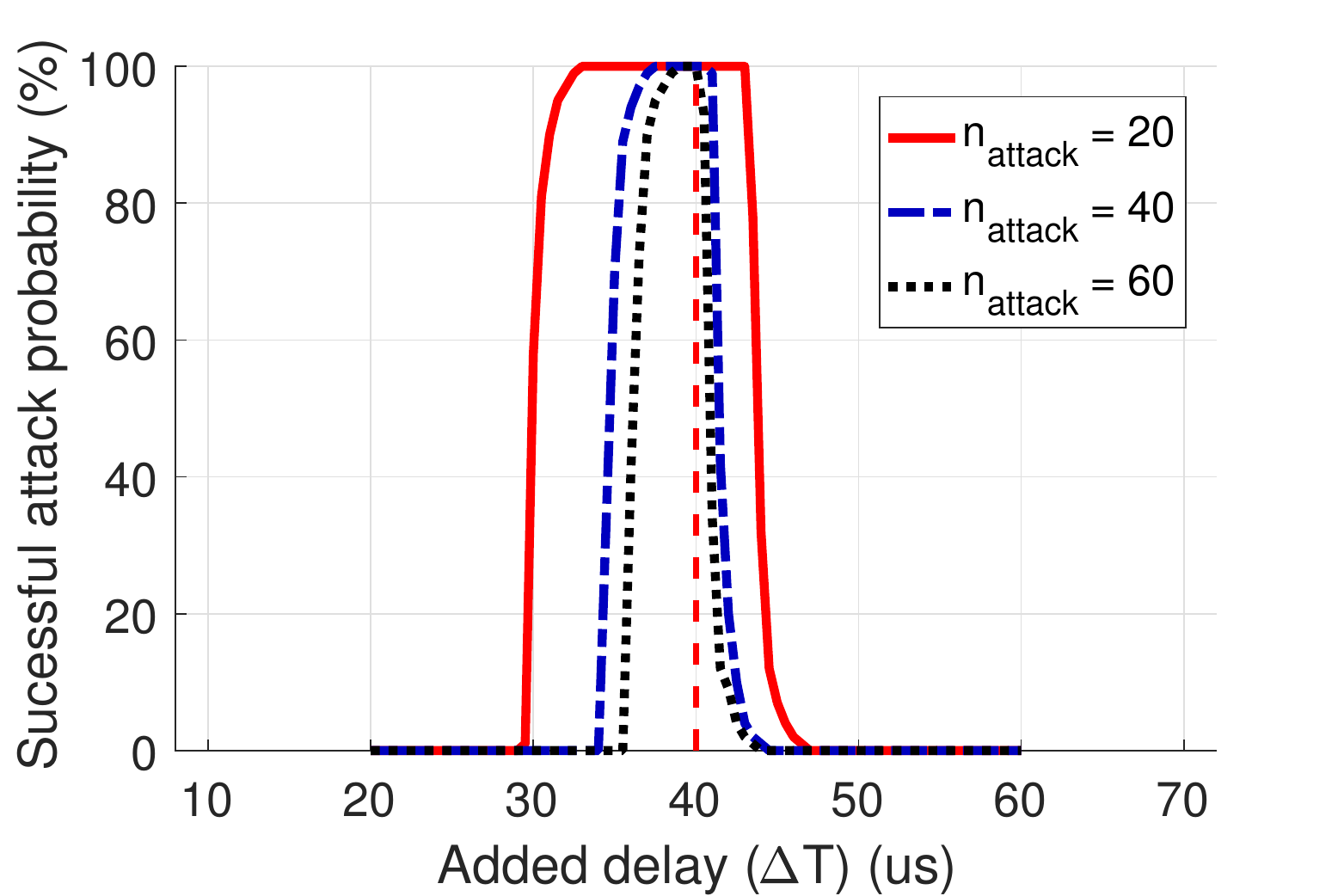}
		\caption{CAN prototype, NTP-based}
		\label{fig:arduino_clock_skew_attack_success_rate_ntp}
	\end{subfigure}
	\begin{subfigure}[h]{0.49\columnwidth} % {0.48\columnwidth}
		\captionsetup{justification=centering}
		\includegraphics[width=\columnwidth]{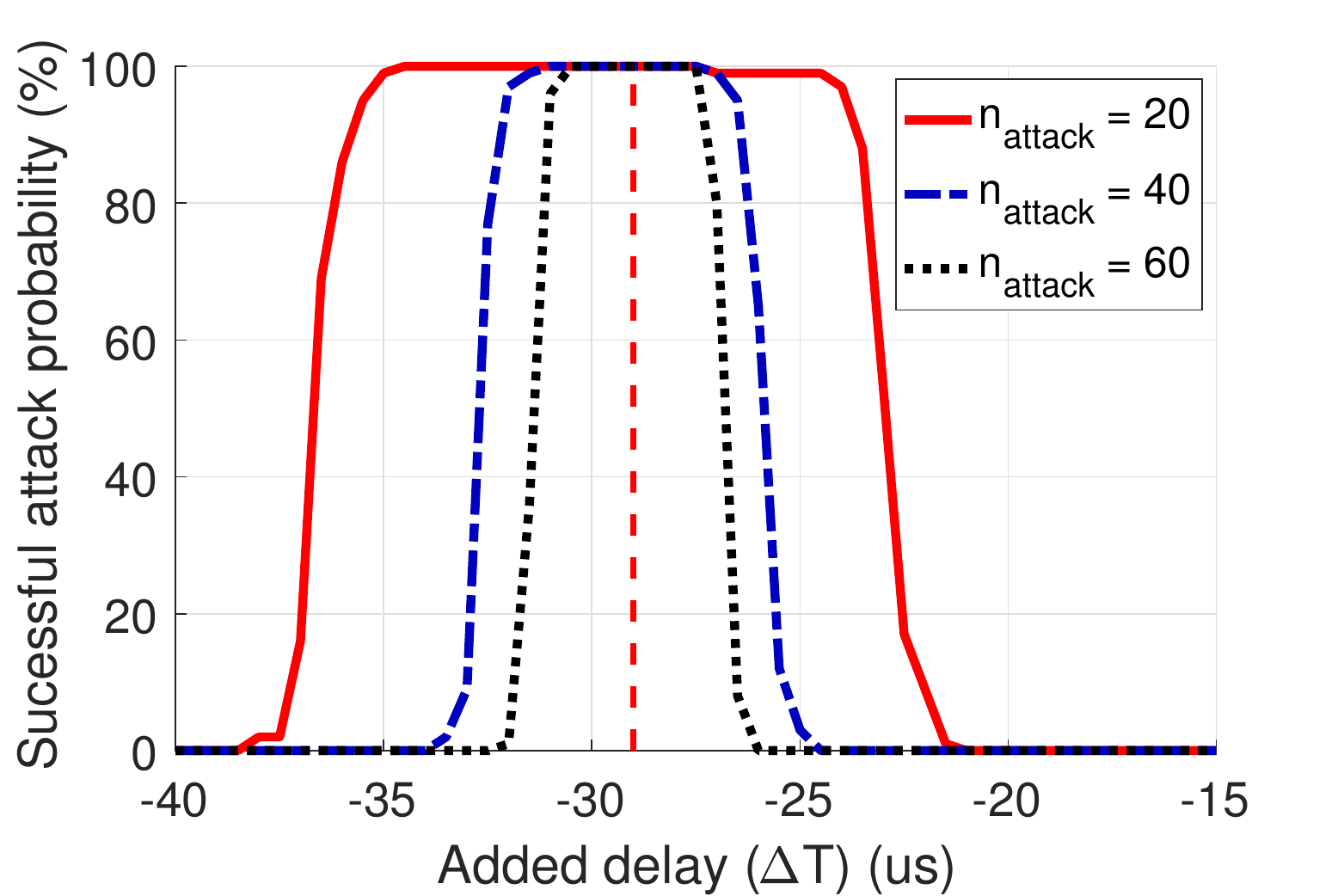}
		\caption{EcoCar testbed, NTP-based}
		\label{fig:ecocar_clock_skew_attack_success_rate_ntp}
	\end{subfigure}
	\caption{Successful attack probability on the  state-of-the-art IDS and the NTP-based IDS on the CAN bus prototype and EcoCar testbed with message period $100$ms. For the value of $\Delta T = 40 \mu s$ achieved in our hardware experiments (red dashed line), the attack was successful in all test cases. The width of each curve is equal to the $\epsilon\text{-MSI}$ for the given detector.
	}
	\label{fig:clock_skew_attack_success_rate}
\end{figure}

%On a CAN testbed, we let the weak attacker, denoted as ECU B, transmit a 10Hz periodic message as the target message, and the strong attacker, denoted as ECU A, launches the cloaking attack on the clock skew detector observing that target message.
%
%In the cloaking attack on the correlation detector, we let 5Hz periodic message of ECU B as the target message that ECU A transmits the attack message based on reception of the target message.
%
%We verify not only the cloaking attack on the CAN testbed but also on a real vehicle, and evaluate the performance of the cloaking attack on the real vehicle.
%
%For the validation of the cloaking attack on the real vehicle, we implement the cloaking attack on and collected data from University of Washington (UW) Ecocar \cite{Crain:2014:Ecocar}.

%Let $n_{attack}$ be the number of batches in one experiment.
%
%Then, we collect $100 \times n_{attack} \times N$ messages, so 100 experiments are prepared.
%
%The successful attack probability, which is the performance metric of the cloaking attack, is defined as the ratio of the number of experiments that bypass the IDS to the total number of experiment which is 100.
%

For the value of $\Delta T$ achieved in our evaluation, the probability of successful attack was $1$ against both the NTP-based IDS and the state-of-the-art IDS (Fig. \ref{fig:clock_skew_attack_success_rate}, dashed line). In order to gain additional insight into the performance of each IDS under cloaking attack, we generated additional data sets by adding different values of $\Delta T$ to the message inter-arrival times, and then analyzed the new datasets using both IDSs.

%Fig. \ref{fig:clock_skew_attack_success_rate} illustrates the performance of the cloaking attack as a function of $\Delta T$ with different $n_{attack}$.
%As we can see, the successful attack probability curve in all plots has a bell shape, which implies that the cloaking attack with a $\Delta T$ value too small or too large will fail to bypass the IDS, but there exists a certain range of $\Delta T$ with a high successful attack probability, referred to as the \textit{slackness}. 
%Intuitively, more slackness implies larger freedom for the attacker.

% Data points (x,y) - (\Delta T, P_s):
% - prototype, state-of-the-art
%  		n_attack=20: (28, 96), (50.5, 96)
% - prototype, NTP
%		n_attack=20: (31.5, 95), (43, 100)
%		n_attack=60: (37.5, 95), (40, 100)

On the CAN bus prototype, with $n_{attack}=20$ and $\epsilon=0.05$, %$0.05-MSI(state-of-the-art \, IDS)$
the $\epsilon\text{-MSI}$ value for the state-of-the-art IDS is $22.5\mu$s  (Fig.~\ref{fig:arduino_clock_skew_attack_success_rate_Cho}), but only $11.5\mu$s for the NTP-based IDS (Fig.~\ref{fig:arduino_clock_skew_attack_success_rate_ntp}).
%
%the slackness is \cmt{[?,?]} with $P_s=95\%$ for the state-of-the-art IDS (Fig.~\ref{fig:arduino_clock_skew_attack_success_rate_Cho}), but only \cmt{[?,?]} for the NTP-based IDS (Fig.~\ref{fig:ecocar_clock_skew_attack_success_rate_Cho}). 
Hence, it is much easier for the cloaking attack to bypass the state-of-the-art IDS than the NTP-based IDS. We also found that increasing $n_{attack}$ has little impact on $\epsilon\text{-MSI}$ for the state-of-the-art IDS, which is $20.5\mu$s for $n_{attack}=40$ or $60$, but significantly impacts $\epsilon\text{-MSI}$ of the NTP-based IDS, which varies from $11.5\mu$s to $2.5\mu$s as $n_{attack}$ is increased from $20$ to $60$. This result suggests that the performance of the NTP-based IDS improves over the attack duration.
Another interesting observation is that the $P_s$ curves are skewed instead of symmetric. 
This is because when the Arduino-based ECU starts operating, its clock skew slowly decreases due to the temperature change in hardware. As a result, the IDS tends to overestimate the clock skew, and is more sensitive to a larger positive delay (that would further decrease the clock skew). 

%Another interesting observation is that the best value of $\Delta T$ (center of $P_s$ curves) seems to be around $39\mu$s  instead of $40\mu$s, as predicted by Eq. (\ref{eq:delta_T}). 
%This is due to quantization errors at both the attacker and the receiving ECU because of Arduino's $4\mu$s resolution, in which case the attacker's choice of $\Delta T$ is less accurate and not optimal, but still effective as a guideline.

%Besides, increasing $n_{attack}$ has little impact on the attack slackness for the state-of-the-art IDS, but would reduce the attack slackness for the NTP-based IDS. 
%For example, with $n_{attack}$ changed from $20$ to $60$, the slackness is changed from \cmt{[?,?]} to \cmt{[?,?]}, and tends to be stable afterwards.

%
%The axis of Fig. \ref{fig:clock_skew_attack_success_rate} is the normalized added delay which is defined as $\frac{\Delta T}{T}$.
%
%The periods of the target messages for the CAN testbed and Ecocar are 100ms and 10ms, respectively.
%

$\epsilon\text{-MSI}$ for the state-of-the-art IDS increases significantly for a real vehicle, as shown in Fig.~\ref{fig:ecocar_clock_skew_attack_success_rate_Cho}, due to the significantly heavier CAN traffic compared to the prototype, which reduces the effectiveness of the detection. 
%When it comes to a real vehicle that has much heavier CAN traffic than the prototype, the slackness for the state-of-the-art IDS increases significantly as shown in Fig.~\ref{fig:ecocar_clock_skew_attack_success_rate_Cho}. 
As an example, a cloaking attack with $\Delta T$ between $-1029\mu$s and $1021\mu$s can bypass the state-of-the-art IDS with $100$\% probability regardless of $n_{attack}$.
For the NTP-based IDS with $\epsilon=0.01$, $\epsilon\text{-MSI}$ is $10.5\mu$s for $n_{attack}=20$, and $3\mu$s for $n_{attack}=60$. Hence, in the real vehicle, as in the CAN prototype, the NTP-based IDS is more effective in detecting masquerade attacks than the state-of-the-art IDS.
The proposed cloaking attack, however, is still able to thwart both detection schemes when $\Delta T$ is chosen to be within the interval $[(\Delta T)_{min}(\epsilon), (\Delta T)_{max}(\epsilon)]$.

\subsection{Performance of Cloaking Attack on Correlation Detector}\label{sec:evaluation_attack_on_correlation}

\begin{figure}[t!]
	\centering
	\begin{subfigure}[h]{0.49\columnwidth} % {0.48\columnwidth}
		\captionsetup{justification=centering}
		\includegraphics[width=\columnwidth]{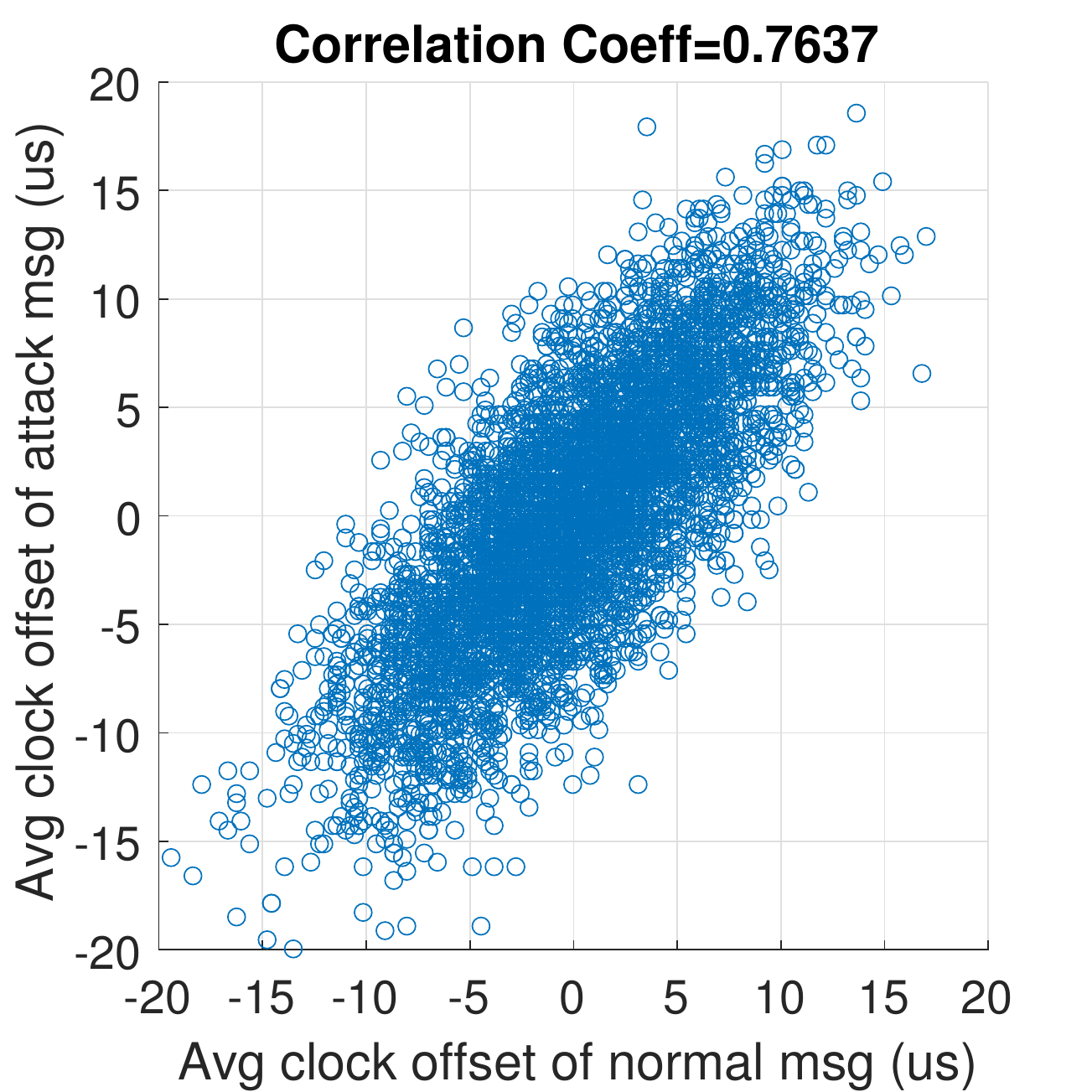}
		\caption{CAN prototype, state-of-the-art}
		\label{fig:arduino_correlation_attack_example_shin}
	\end{subfigure}
	\begin{subfigure}[h]{0.49\columnwidth} % {0.48\columnwidth}
		\captionsetup{justification=centering}
		\includegraphics[width=\columnwidth]{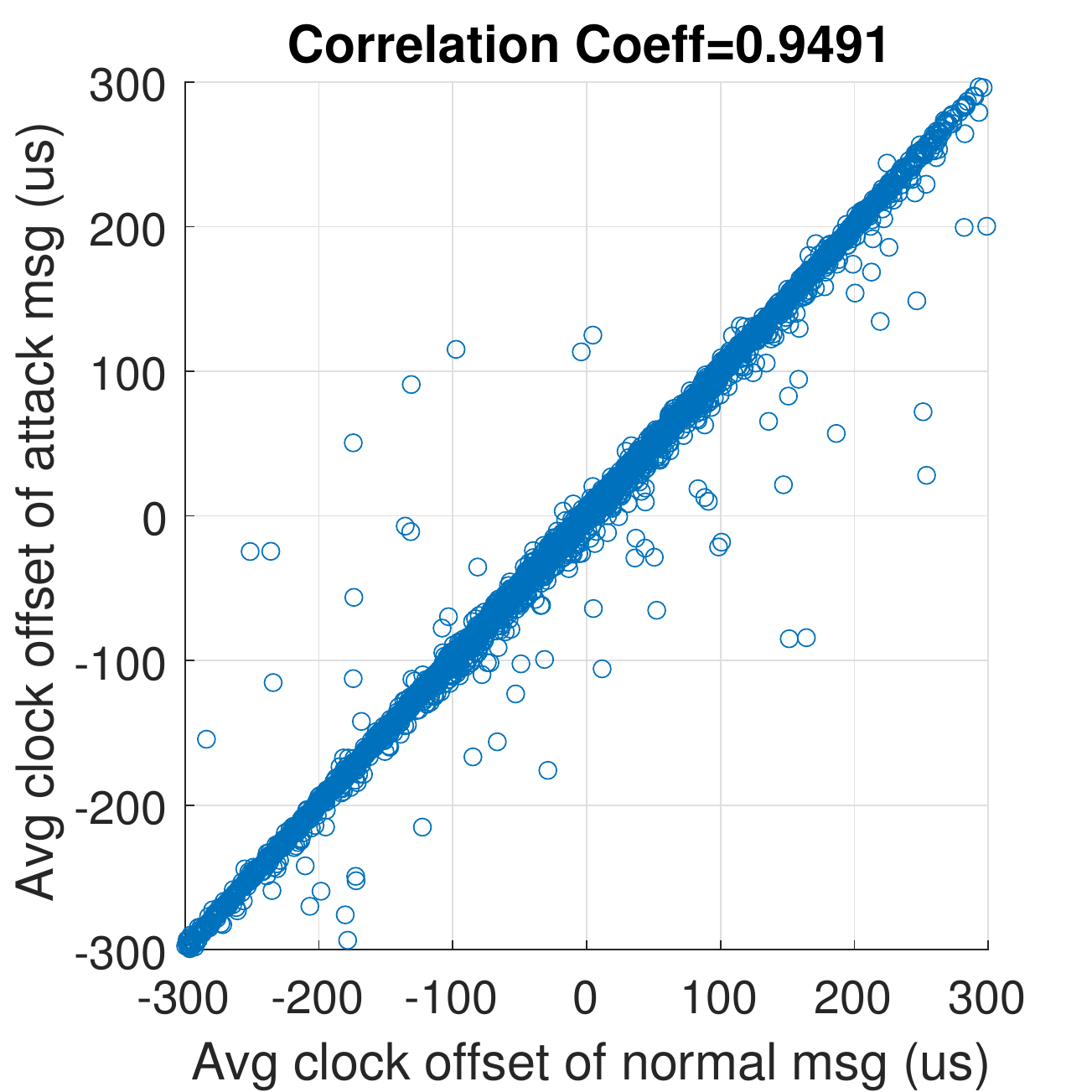}
		\caption{EcoCar testbed, state-of-the-art}
		\label{fig:ecocar_correlation_attack_example_shin}
	\end{subfigure}
	\\
	\begin{subfigure}[h]{0.49\columnwidth} % {0.48\columnwidth}
		\captionsetup{justification=centering}
		\includegraphics[width=\columnwidth]{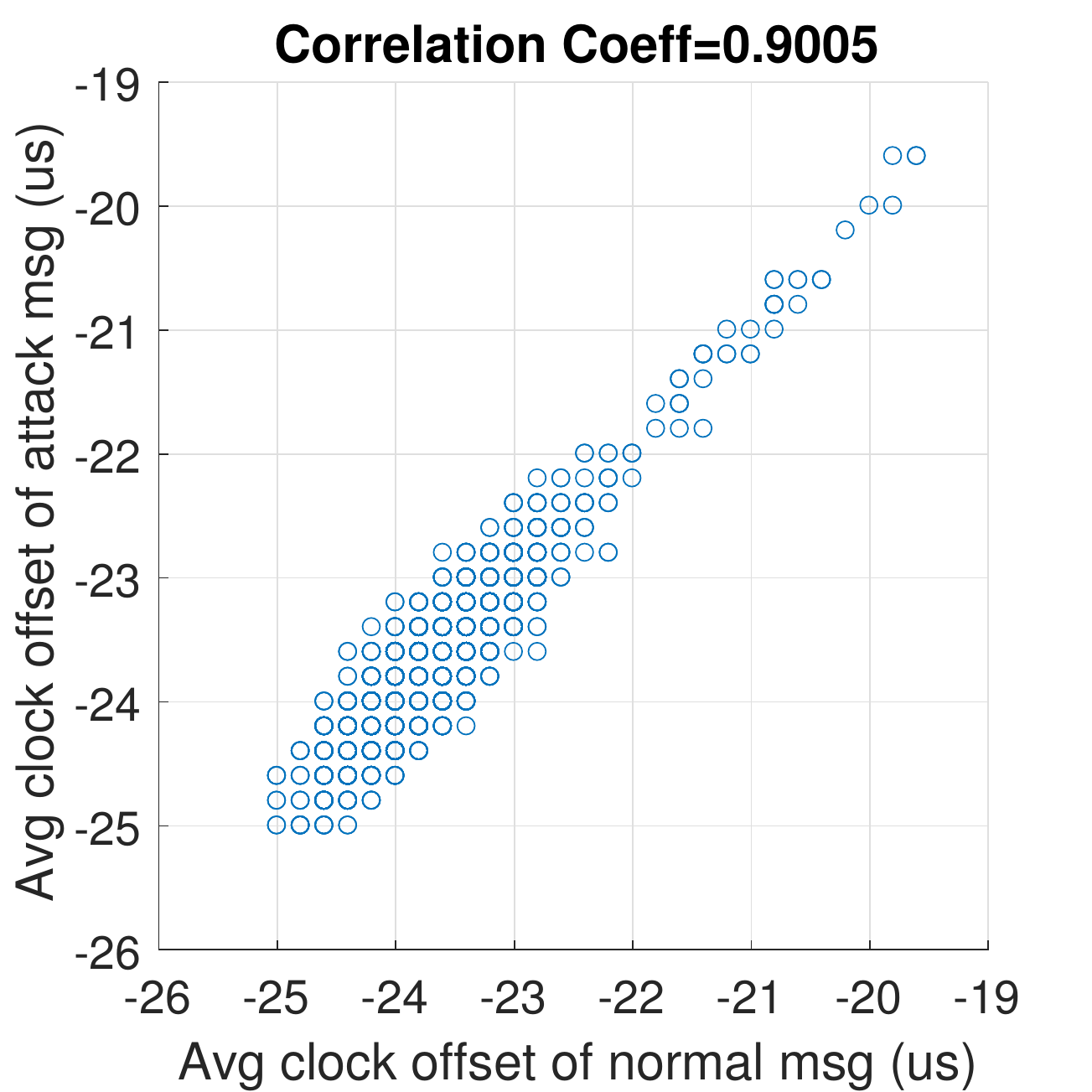}
		\caption{CAN prototype, NTP-based}
		\label{fig:arduino_correlation_attack_example_ntp}
	\end{subfigure}
	\begin{subfigure}[h]{0.49\columnwidth} % {0.48\columnwidth}
		\captionsetup{justification=centering}
		\includegraphics[width=\columnwidth]{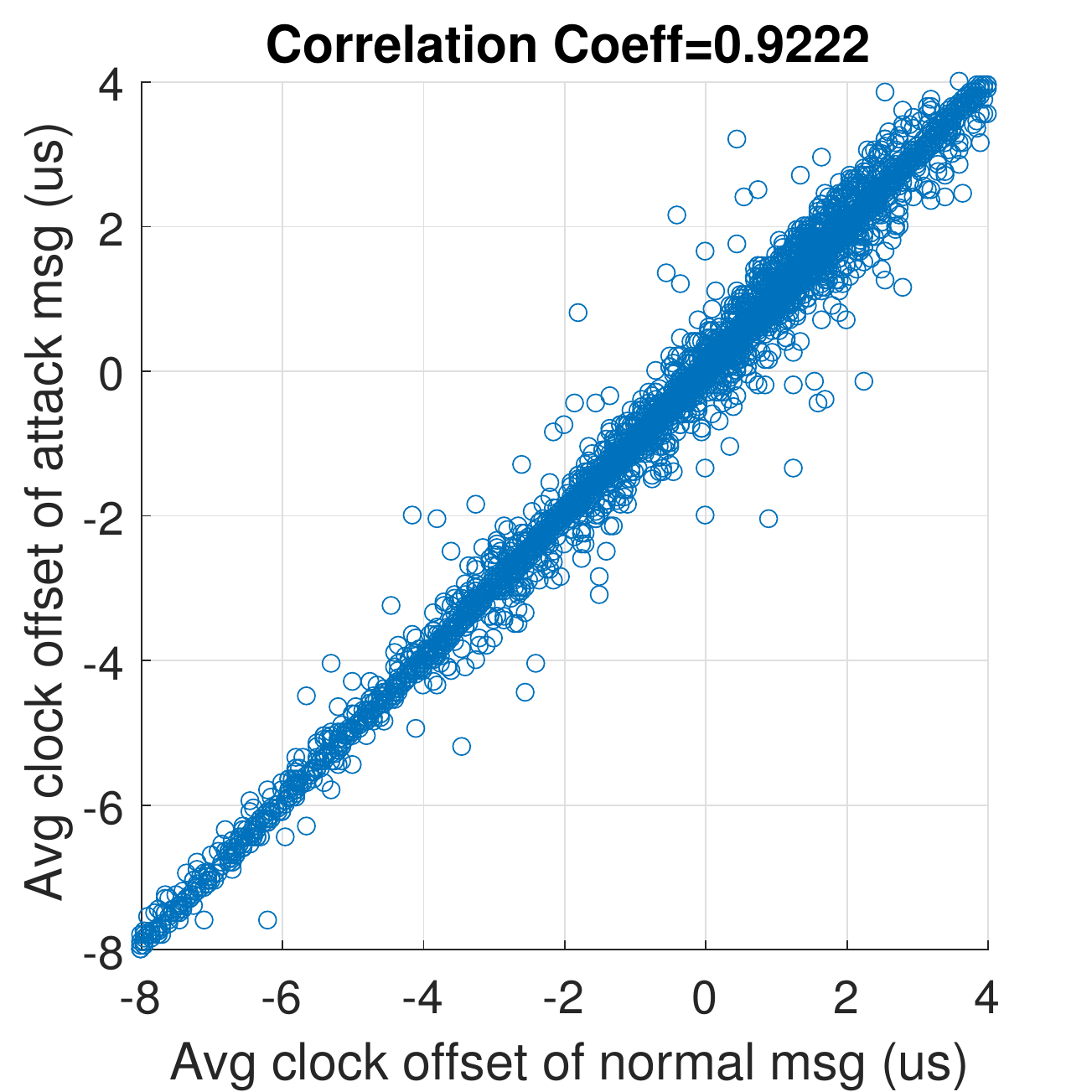}
		\caption{EcoCar testbed, NTP-based}
		\label{fig:ecocar_correlation_attack_example_ntp}
	\end{subfigure}
	\caption{Scatter plot of the average offsets of sibling messages and attack messages under cloaking attack. The correlation coefficient is above 0.9 in the EcoCar testbed, and hence is comparable to the coefficient for consecutive messages.}
	\label{fig:correlation_attack_example}
\end{figure}

In this section, we demonstrate and evaluate the cloaking attack on the correlation detector.
On the CAN bus prototype, the targeted message is $5$Hz.
When launching the cloaking attack, the Arduino-based impersonating ECU transmits a spoofed message 0x11 after it observes a sibling message of the targeted message, with a constant delay of $100$ms\footnote{
		As mentioned in Section~\ref{sec:cloaking_on_correlation_analysis}, as long as two messages are received with a constant delay, they will be highly correlated. To validate this, we programmed the strong attacker to transmit after a constant delay instead of immediately on the CAN bus prototype.}.
On the EcoCar testbed, two $100$Hz messages 0xC1 and 0xC5 from a stock ECU are identified to be highly correlated.
We choose 0xC5 as the target, and 0xC1 as its sibling message.
Due to limited computing capabilities, the Arduino-based ECU is not able to receive all messages on the CAN bus, filter for the sibling message, and transmit the spoofed message. 
Hence, we use the Raspberry-Pi-based ECU as the impersonating ECU.
It injects messages with a non-conflicting ID 0xC0, instead of 0xC5, in order to avoid any undesirable impact on the EcoCar.
A total of $14$ hours and $1.2$ hours of attack data were collected from the CAN bus prototype and the EcoCar testbed, respectively.
As a baseline, we collected $4.7$ hours of normal data with one ECU transmitting two messages consecutively on the CAN bus prototype. 
For the EcoCar testbed, since the targeted message is not suspended (for safety), the data we collected also contains the normal data. 
The same settings in Section~\ref{sec:evaluation_attack_on_clock_skew} are used for state-of-the-art and NTP-based IDSs.

Fig.~\ref{fig:correlation_attack_example} shows a typical scatter plot of average offsets of the sibling message and the attack message, when the cloaking attack is mounted. 
For the CAN bus prototype, the correlation is $0.76$ and $0.90$ for state-of-the-art and NTP-based IDSs, respectively.
This is mainly because an Arduino-based ECU is dedicated to transmission, which implies a smaller jitter and offset deviation, while the quantization error is quite significant due to the Arduino's $4\mu$s time resolution.
On the EcoCar testbed, the cloaking attack can achieve correlation up to $0.95$ and $0.92$ for state-of-the-art and NTP-based IDSs.

\begin{figure}[ht!]
	\centering
	\includegraphics[width=0.65\columnwidth]{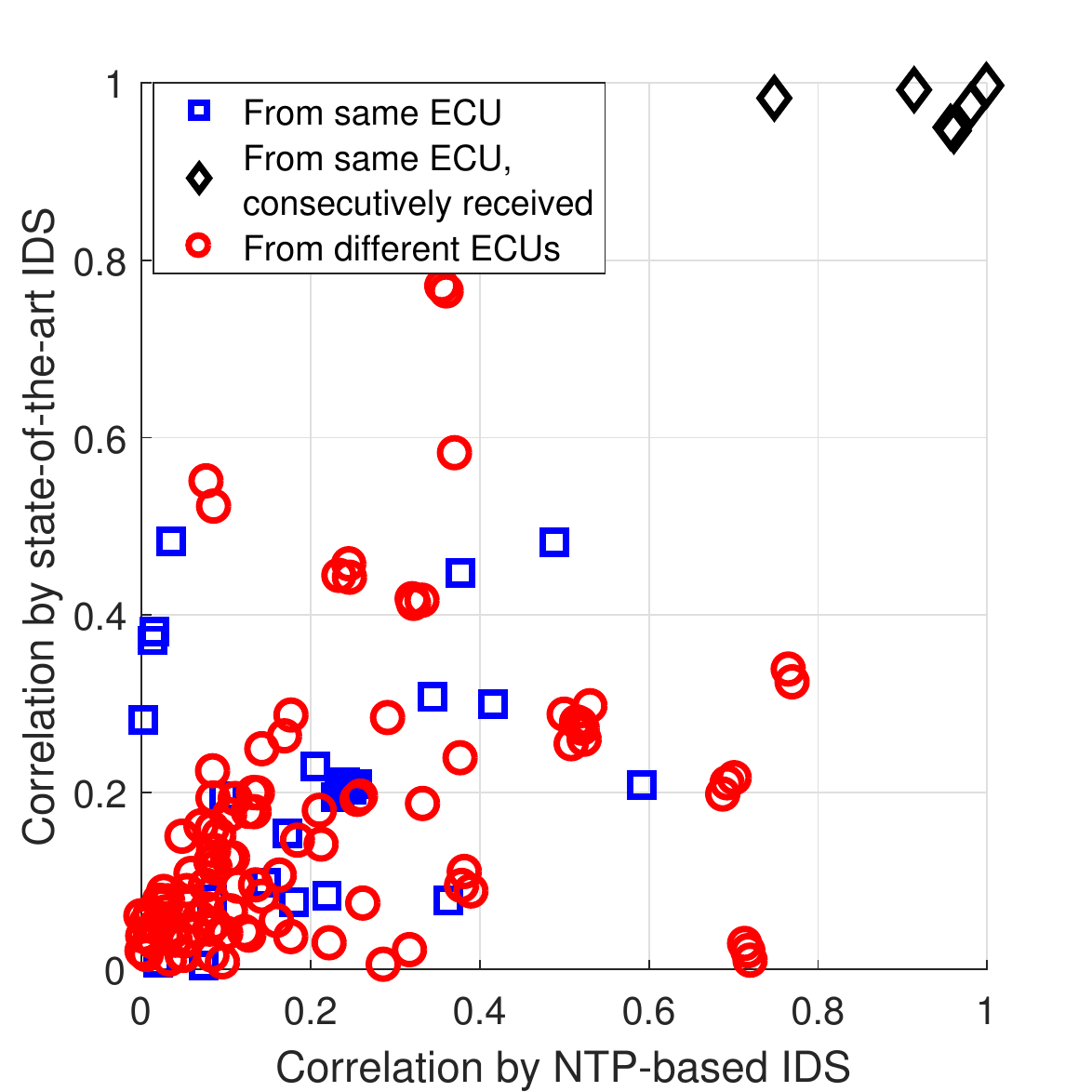}
	\caption{Correlation relationship between pairwise messages on the EcoCar testbed. 
	Consecutive messages from the same ECU are highly correlated, while others are less correlated.}
	\label{fig:correlation_analysis_pairs}
\end{figure}

To understand the correlation relationship between pairwise messages on the EcoCar testbed, we examine $17$ messages from $5$ ECUs with periods of $10$ms, $12$ms or $100$ms, based on the ground truth provided by the manufacturer. 
All pairs of messages are classified into the following three categories: 1) from the same ECU and (almost always) received consecutively, 2) from the same ECU but not received consecutively, or 3) from different ECUs.
Correlation values are computed using 200 batches.
As illustrated in Fig.~\ref{fig:correlation_analysis_pairs}, for two messages from different ECUs, their correlation is generally low (e.g., less than $0.2$) for both state-of-the-art and NTP-based IDSs.
In addition, not all pairs of messages from the same ECU have high correlation: $81$\% of them have correlation less than $0.6$, and there are only 5 pairs with correlation higher than $0.9$ for both IDSs.
We checked such pairs and confirmed that their messages are always consecutively received.
This result is indeed consistent with our analysis in Section~\ref{sec:cloaking_on_correlation_analysis}.

\begin{figure}[h]
	\centering
	\begin{subfigure}[h]{0.49\columnwidth} % {0.48\columnwidth}
		\captionsetup{justification=centering}
		\includegraphics[width=\columnwidth]{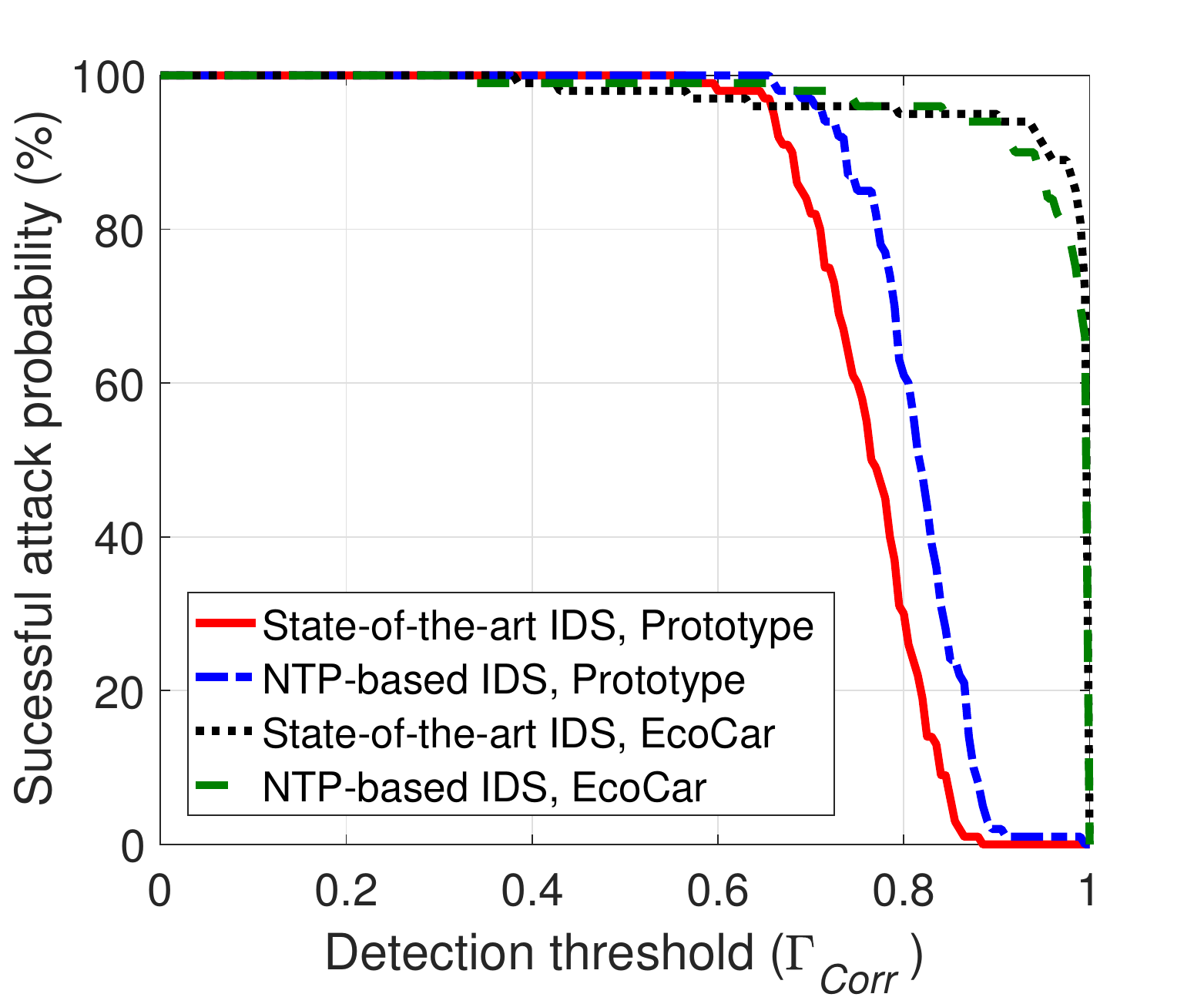}
		\caption{Successful attack probability}
	\end{subfigure}
	\begin{subfigure}[h]{0.49\columnwidth} % {0.48\columnwidth}
		\captionsetup{justification=centering}
		\includegraphics[width=\columnwidth]{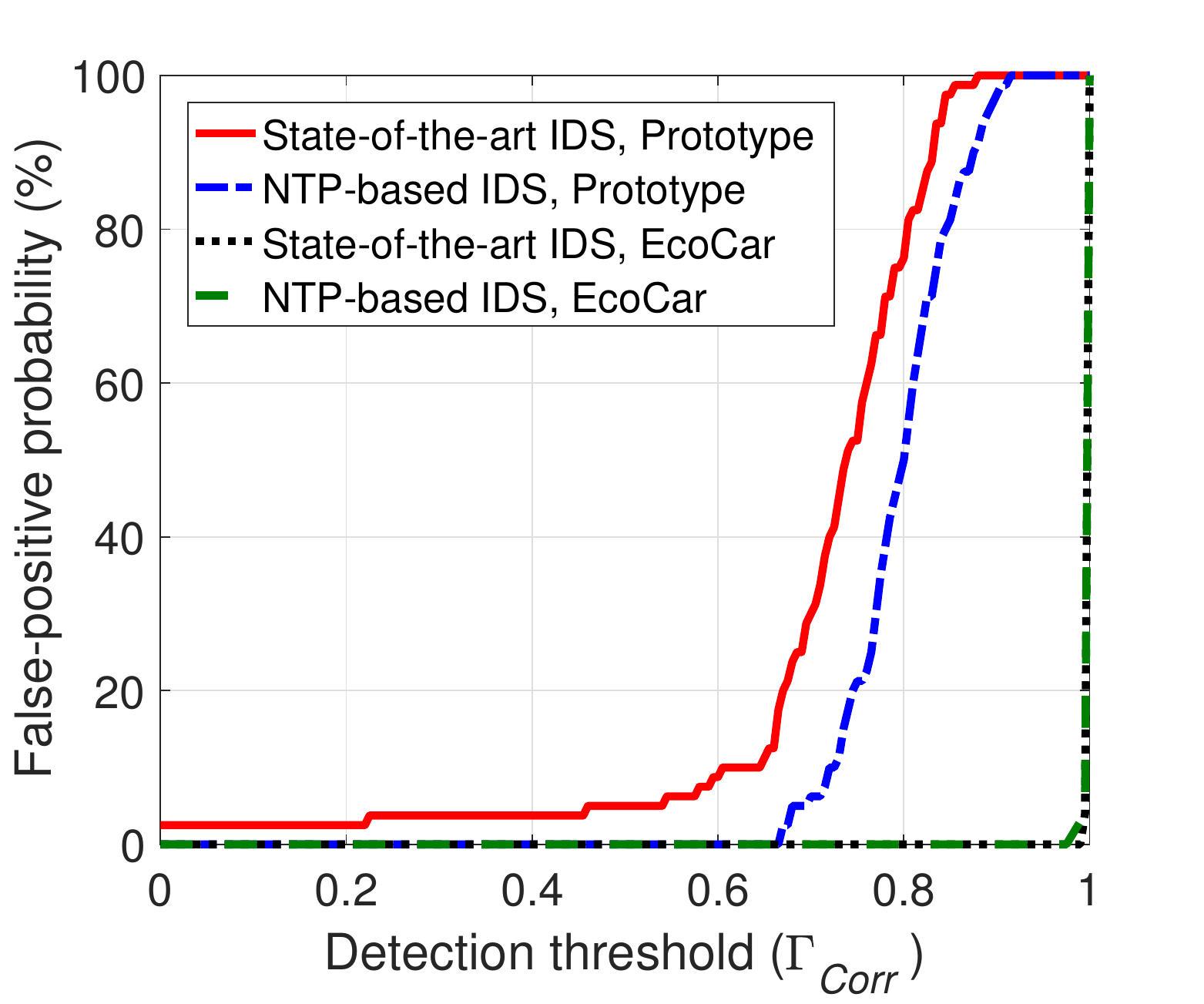}
		\caption{False alarm probability}
	\end{subfigure}
	\caption{Successful attack probability and false-positive probability of the cloaking attack on the correlation detector under changing detection threshold. In the CAN prototype, if the detector is chosen to achieve probability of false alarm $P_{fa} \leq 0.05$, then the attack succeeds with probability at least 0.95. In the EcoCar, the probability of success for the attack is 0.8 when the detector parameters are chosen so that $P_{fa} = 0$.
}
	\label{fig:cloaking_attack_on_correlation}
\end{figure}

Next we evaluate the performance of the cloaking attack. 
An attack on the correlation detector is successful if the resulting correlation is higher than or equal to the detection threshold $\Gamma_{corr}$, and failed otherwise.
A total of $100$ experiments using the attack data are conducted, each consisting of $50$ batches, to compute the successful attack probability $P_s$.
Intuitively, a higher $\Gamma_{corr}$ may cause a IDS to report a false alarm, i.e., declaring an attack when there is actually none. 
The false alarm probability $P_{fa}$ is equal to the percentage of experiments where the IDS reports a false alarm.
A total of $80$ experiments using the normal data are conducted to compute $P_{fa}$.

%
%In addition to the CAN testbed, the cloaking attack on the correlation detector is mounted on Ecocar, and the clock offset of the attack message can be highly correlated using the cloaking attack on the correlation detector as shown in Fig. \ref{fig:ecocar_correlation_attack_example}.
%
%(In \cite{Shin:2016:finger}, IDS decides the two messages are transmitted from one ECU if the correlation coefficient is greater than 0.8.)

%Also, the cloaking attack on the correlation detector matches the strong attacker's clock skew to the weak attacker's clock skew as shown in \hl{Fig.}.
%
%Hence, the cloaking attack on the correlation detector is not caught by the clock skew detector as well.
%\hl{(In case, I can provide sets of figures which are avg offset, acc offset, and control limits to rigorously show that cloaking attack on correlation detector is not detected by clock skew detector.)}

%Then, we evaluate the performance of the cloaking attack on the correlation detector with respect to the successful attack probability.
%
%Each experiment is consist of 50 batches and 100 experiments are considered.
%

Fig.~\ref{fig:cloaking_attack_on_correlation} illustrates $P_s$ and $P_{fa}$ as a function of the detection threshold $\Gamma_{corr}$. 
As we can see, a larger $\Gamma_{corr}$ decreases $P_s$, making the attack more difficult, but also leads to more false alarms.
On the CAN bus prototype, the state-of-the-art IDS needs to set $\Gamma_{corr}$ to 0.54 to ensure $P_{fa}\leq$5\%, at which point we have $P_s=$100\%.
For the NTP-based IDS, when $\Gamma_{corr}$ is 0.68, we have $P_{fa}\leq$5\% and $P_s=$98\%.
It demonstrates that the cloaking attack is able to effectively bypass both IDSs. 

On the EcoCar testbed, when $\Gamma_{corr}$ is $0.8$, $P_s$ is 95\% and 96\% for state-of-the-art and NTP-based IDS, respectively. 
Since the two messages have very high correlation under the normal condition, $\Gamma_{corr}$ may be set to 0.975 without any false alarm.
At this point, $P_s$ for state-of-the-art and NTP-based IDSs is 89\% and 80\%, respectively. 
It is important to note that such attack performance is already achieved with a lower-end ECU based on Raspberry Pi, and we would expect $P_s$ to increase when the cloaking attack is mounted by the strong attacker inside a vehicle, which is left as our future work.

%>>> SECTION: Discussion
% \input{./sections/discussion}

%>>> SECTION: Conclusion
\section{Conclusion}
\label{sec:conclude}
This paper investigated attacks on in-vehicle networks, in which an adversary compromises one or more ECUs and introduces spoofed messages claiming to be from a targeted ECU. Recent works have proposed using the ECU clock skew as a fingerprint to detect attacks, resulting in clock skew-based intrusion detection systems (IDS) that make use of first- and second-order moments. In this paper, we proposed the cloaking attack on IDS, in which an adversary changes the transmission times of spoofed messages in order to match the clock skew of the targeted ECU. We evaluated the cloaking attack on  a CAN bus prototype and a connected vehicle, and showed that the state-of-the-art IDS was deceived in all test cases. We also proposed and evaluated a novel IDS based on the Network Time Protocol. 
In order to quantify the effectiveness of the attack, we presented a new security metric, the Maximum Slackness Index, which is the range of added delay that the adversary actions can introduce before being detected. This work makes the case that the impact of  coupling between cyber and physical components in CPS security needs to be understood, especially when attempting to leverage physical invariants arising from physical components to provide security assurances.

% the need to understand the coupling between cyber and physical components in CPS security, especially when leveraging physical invariants to detect cyber attacks.

%>>> SECTION: Acknowledgment
% \section*{Acknowledgment}

%>>> SECTION: Reference
\bibliographystyle{IEEEtran}
\bibliography{./sections/sang_bib}

%\section*{References}
%\begin{thebibliography}{00}
%\bibitem{b1} G. Eason, B. Noble, and I. N. Sneddon, ``On certain integrals of Lipschitz-Hankel type involving products of Bessel functions,'' Phil. Trans. Roy. Soc. London, vol. A247, pp. 529--551, April 1955.
%\bibitem{b2} J. Clerk Maxwell, A Treatise on Electricity and Magnetism, 3rd ed., vol. 2. Oxford: Clarendon, 1892, pp.68--73.
%\bibitem{b3} I. S. Jacobs and C. P. Bean, ``Fine particles, thin films and exchange anisotropy,'' in Magnetism, vol. III, G. T. Rado and H. Suhl, Eds. New York: Academic, 1963, pp. 271--350.
%\bibitem{b4} K. Elissa, ``Title of paper if known,'' unpublished.
%\bibitem{b5} R. Nicole, ``Title of paper with only first word capitalized,'' J. Name Stand. Abbrev., in press.
%\bibitem{b6} Y. Yorozu, M. Hirano, K. Oka, and Y. Tagawa, ``Electron spectroscopy studies on magneto-optical media and plastic substrate interface,'' IEEE Transl. J. Magn. Japan, vol. 2, pp. 740--741, August 1987 [Digests 9th Annual Conf. Magnetics Japan, p. 301, 1982].
%\bibitem{b7} M. Young, The Technical Writer's Handbook. Mill Valley, CA: University Science, 1989.
%\end{thebibliography}

%>>> SECTION: Appendix
\newpage
\appendix
%Appendix A
\subsection{Workflow of IDS}\label{appendix:workflow}

\begin{figure}[ht!]
	\centering
	\includegraphics[width=1\columnwidth]{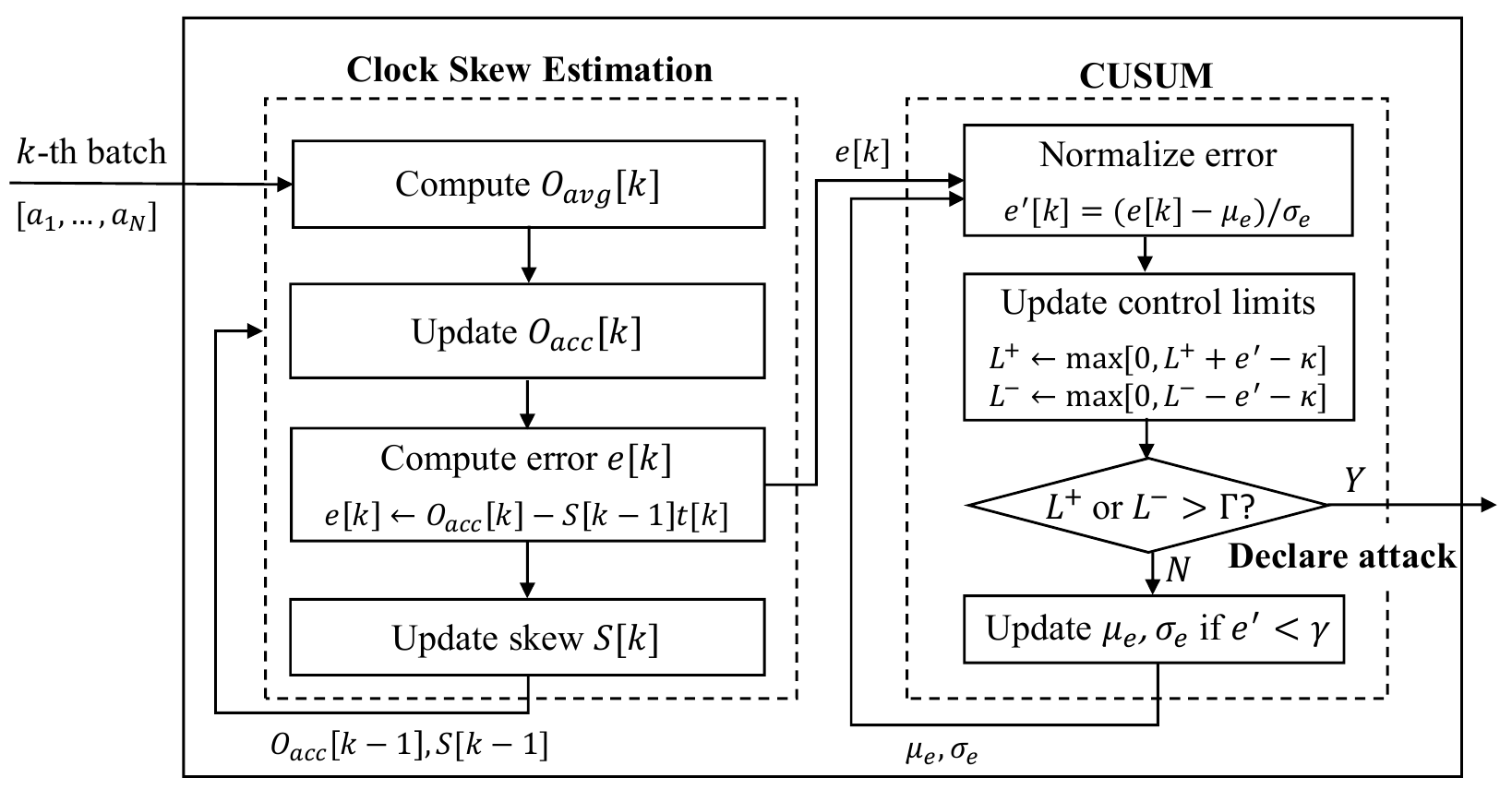}
	\caption{Workflow of the state-of-the-art IDS \cite{Shin:2016:finger} and the NTP-based IDS.
		There are two main blocks: clock skew estimation and CUSUM.
		The clock skew estimation block updates the clock skew using the most recent batch. 
	 	CUSUM takes the identification error and update control limits for detection. 	
}
	\label{fig:cho_shin_cids_flow}
\end{figure}

Fig. \ref{fig:cho_shin_cids_flow} describes an IDS workflow. The following steps are applicable to both the state-of-the-art IDS in \cite{Shin:2016:finger} and the NTP-based IDS that we propose.
An IDS consists of two blocks: clock skew estimation and Cumulative Sum (CUSUM). 
The clock skew estimation block takes the timestamps of $N$ newly arrived messages as input.
For the $k$-th batch, the state-of-the-art IDS computes the average offset $O_{avg}[k]$ and the accumulated offset $O_{acc}[k]$ according to Eq.~(\ref{eq:cho_shin_avg_offset}) and Eq.~(\ref{eq:cho_shin_acc_offset}), respectively, whereas the NTP-based IDS follows Eq.~(\ref{eq:ntp_avg_offset}) and Eq.~(\ref{eq:ntp_acc_offset}).
%Then the identification error is computed, and fed into the Recursive Least Square for clock skew tracking.
Then the identification error $e[k]$ is computed, and used to obtain the updated clock skew $S[k]$ using the Recursive Least Square algorithm \cite{Shin:2016:finger,Haykin:2011}.

%
%Clock skew detector of the IDS is composed of two blocks: clock skew estimation and CUSUM.
%
%In the clock skew estimation block, the IDS computes clock skew of ECU using periodic messages.
%
%Let a batch is defined as a set of $N$ newly arrived messages.
%
%The state-of-the-art IDS first computes the average offset for every $N$ newly arrived messages from the receive timestamp of the first message of that batch.
%
%On the other hand, the NTP-based IDS computes the average offset using the last messages of the previous batch and the current batch.
%
%Then, accumulated offset is calculated by adding the absolute value of the average offset of each batch or the original value of the average offset of each batch in the state-of-the-art IDS and the NTP-based IDS, respectively.
%
%Since the clock skew is assumed to be constant, the clock skew is estimated by using the recursive least square estimation method which finds the clock skew that minimizes the identification error \cite{Shin:2016:finger}.
%
%The current estimated clock skew and accumulated offset is fed back to the next batch to estimate clock skew.

%\rev{
The CUSUM block takes the identification error as input.
It starts to maintain the statistics of all past identification errors, i.e., mean and standard deviation, after $n_{init}$ (e.g., $50$) error samples are received. 
Then the new error sample is first normalized, and used to update the control limits. 
If either upper or lower control limit ($L^+/L^-$) exceeds the detection threshold $\Gamma$, the IDS declares an attack. 
In order to be robust against noise, if the normalized error $e'[k]$ is less than the threshold $\gamma$, the error statistics will be updated using the new error sample $e[k]$ and all past error samples; otherwise, $e[k]$ will be dropped and error statistics will not be updated.
In practice, online algorithms like Welford algorithm \cite{welford:1962} may be used to update error statistics.

\subsection{Impact of Mistiming on Cloaking Attack}\label{appendix:mistiming}
In a masquerade or cloaking attack, the strong attacker needs to start transmitting the spoofed message at the time constant at which the targeted message should have been transmitted, if it had not been suspended. 
It naturally raises the question whether mistiming affects the cloaking attack performance.
In this simulation, we introduce a mistiming delay (either positive or negative) between the last message of normal data and the first message of attack data, in addition to the message period.
%When the cloaking attack initiate, the strong attack cannot start transmitting the attack message exactly after the period of the target message.
%
%We define the mistiming as the period of the target message minus the difference between the receive timestamps of the last target message and the first attack message.
%
%If the mistiming is negative, the attack message is injected earlier than the period of the message.

The IDS is fed with $1000$ batches of normal data, followed by $n_{attack}$ batches of attack data with a batch size of $20$ in each experiment.
$\Gamma$ is 5 for both IDSs, and $\gamma$ is set to $3$ and $4$ for state-of-the-art and NTP-based IDSs, respectively.
Also, $\kappa$ is set to $5$ for the CAN bus prototype and to $8$ for the EcoCar testbed, respectively.

\begin{figure}[t!]
	\centering
	\begin{subfigure}[h]{0.49\columnwidth} % {0.48\columnwidth}
		\captionsetup{justification=centering}
		\includegraphics[width=\columnwidth]{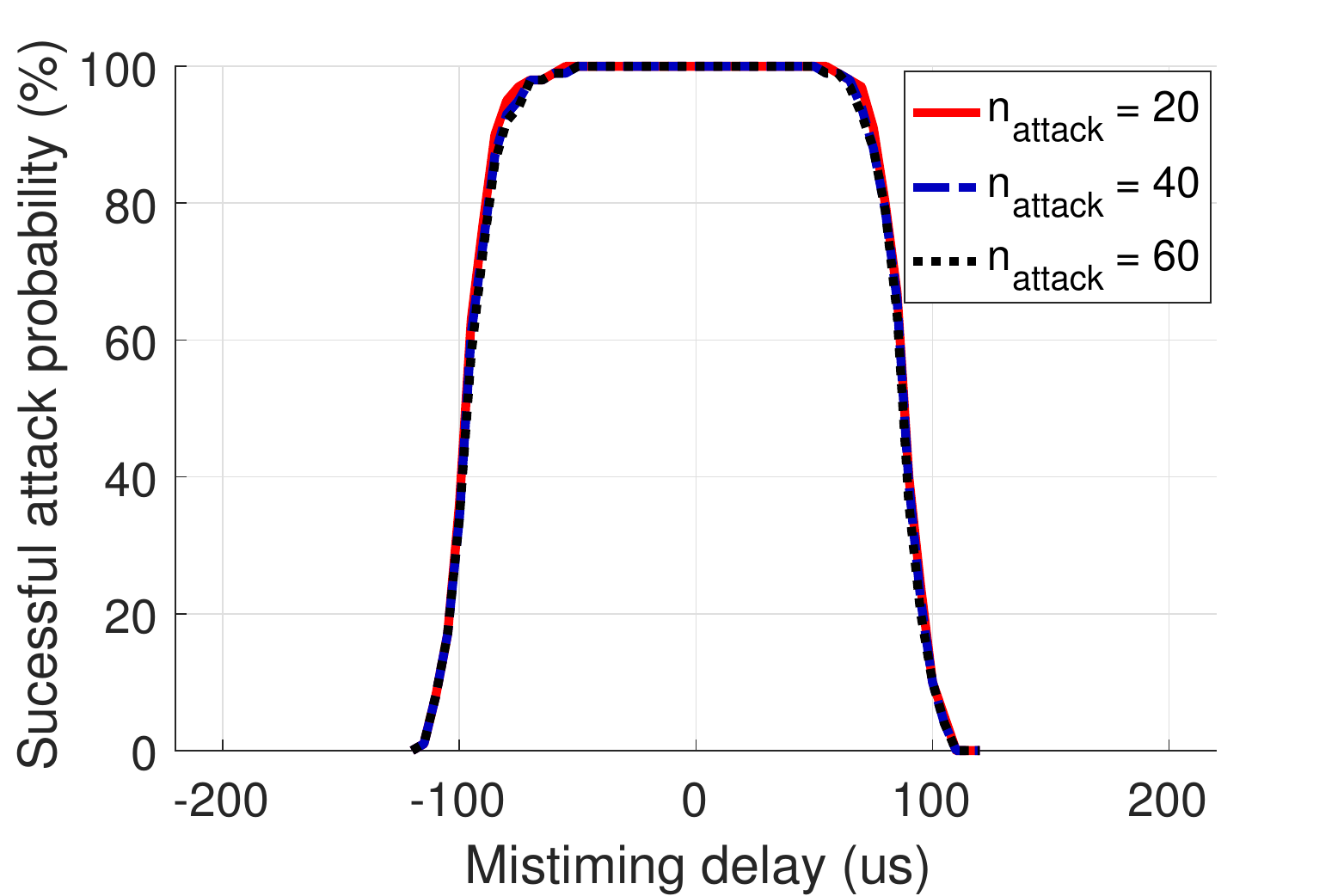}
		\caption{CAN prototype, state-of-the-art}
		\label{fig:arduino_clock_skew_attack_success_rate_Cho_mistime}
	\end{subfigure}
	\begin{subfigure}[h]{0.49\columnwidth} % {0.48\columnwidth}
		\captionsetup{justification=centering}
		\includegraphics[width=\columnwidth]{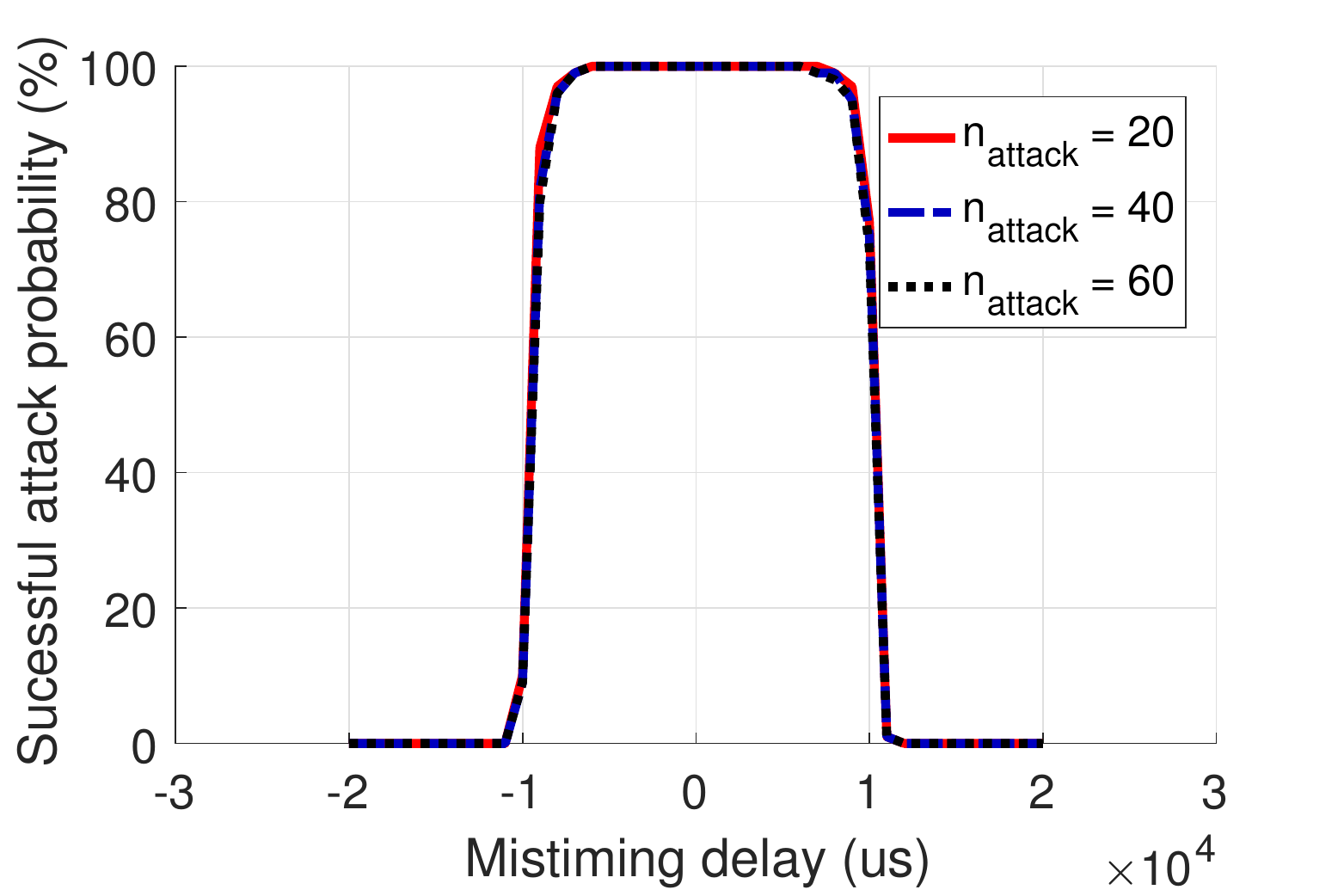}
		\caption{EcoCar testbed, state-of-the-art}
		\label{fig:ecocar_clock_skew_attack_success_rate_Cho_mistime}
	\end{subfigure}
	\\
	\begin{subfigure}[h]{0.49\columnwidth} % {0.48\columnwidth}
		\captionsetup{justification=centering}
		\includegraphics[width=\columnwidth]{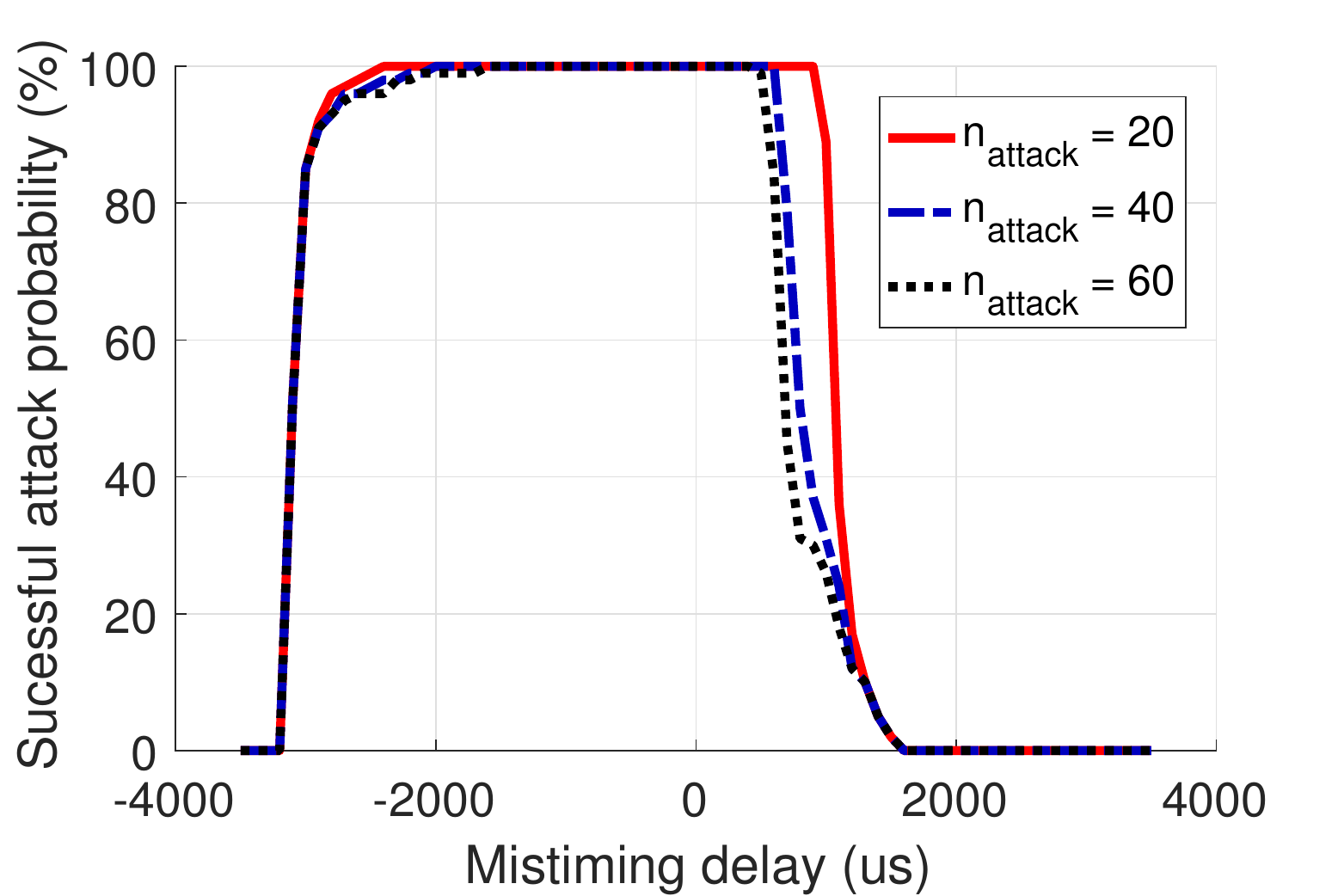}
		\caption{CAN prototype, NTP-based}
		\label{fig:arduino_clock_skew_attack_success_rate_ntp_mistime}
	\end{subfigure}
	\begin{subfigure}[h]{0.49\columnwidth} % {0.48\columnwidth}
		\captionsetup{justification=centering}
		\includegraphics[width=\columnwidth]{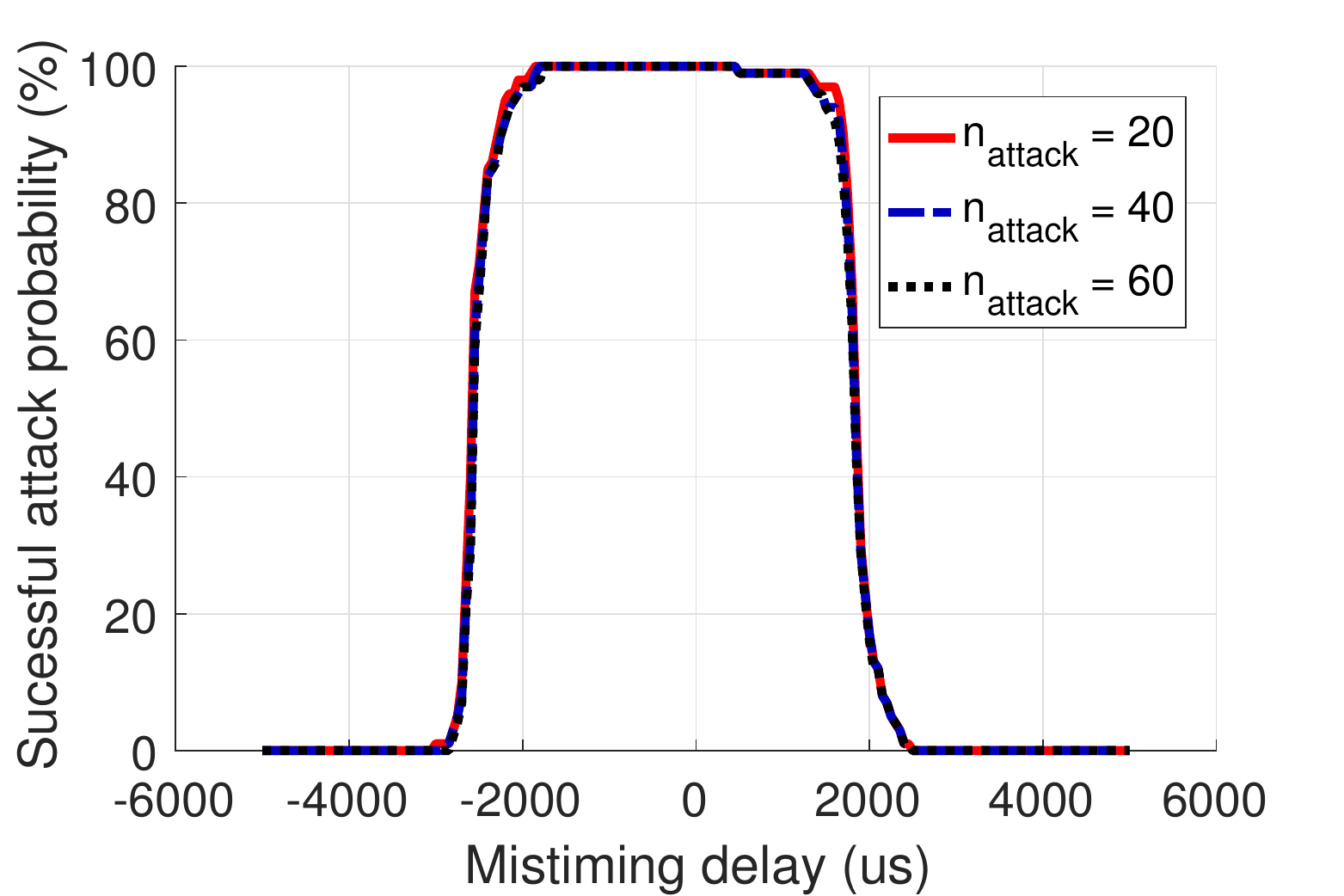}
		\caption{EcoCar testbed, NTP-based}
		\label{fig:ecocar_clock_skew_attack_success_rate_ntp_mistime}
	\end{subfigure}
	\caption{Impact of the mistimed cloaking attack on the state-of-the-art IDS and the NTP-based IDS. If the strong attacker can inject the fist attack message on the proper time, the cloaking attack can bypass both IDSs.
	}
	\label{fig:clock_skew_attack_success_rate_mistime}
\end{figure}

Fig. \ref{fig:clock_skew_attack_success_rate_mistime} shows the impact of the mistiming of the cloaking attack on state-of-the-art and NTP-based IDSs.
In general, larger mistiming causes the attack performance to decrease.
On the CAN bus prototype, any amount of mistiming between $-55\mu$s and $55\mu$s does not affect the attack performance (i.e., $P_s$ is $100\%$ with $n_{attack}=60$) against the state-of-the-art IDS, whereas the allowed mistiming is much larger for the NTP-based IDS, mainly due to the difference in clock skew estimation. 
Since the clock skew of the Arduino-based ECU slowly decreases due to the temperature change in hardware as it warms up, the estimator tends to overestimate the clock skew, and thus is more sensitive to larger positive mistiming (that would further decrease the clock skew), which explains the skewness of the curves in Fig. \ref{fig:arduino_clock_skew_attack_success_rate_ntp_mistime}. 

On the EcoCar tested, the allowed mistiming is increased significantly, which is between $-6$ms to $7$ms for the state-of-the-art IDS, and between $-1.8$ms and $0.5$ms for the NTP-based IDS, due to much heavier CAN traffic in a real vehicle. 
The above observations imply that the timing is hardly a strict requirement for the adversary to launch a clocking attack in a real vehicle.

%
%Compared with the order of the added delay $\Delta T$, the order of the mistiming is $10$ to $10^3$ times higher than $\Delta T$ which indicates that the impact of the mistiming is less significant than $\Delta T$.
%
%When $n_{attakck}$ is 20, the range of the mistiming in which the successful attack probability is 1 for the state-of-the-art IDS is $-55\mu$s to $55\mu$s in the CAN bus prototype whereas the range is $-6000\mu$s to $7000\mu$s in the Ecocar testbed as shown in Figs. \ref{fig:arduino_clock_skew_attack_success_rate_Cho_mistime} and \ref{fig:ecocar_clock_skew_attack_success_rate_Cho_mistime}.
%  
%That is because the periodicity of the messages deviates much larger in the Ecocar testbed than the CAN bus prototype since lots of messages are transmitted with the high frequency in the EcoCar testbed.

% Successful Rates

\end{document}